\documentclass[twocolumn,notitlepage,superscriptaddress, nofootinbib,nobibnotes, aps,prd,10pt]{revtex4-1}
\usepackage{amssymb, fp}
\usepackage{graphicx}
\usepackage{amsmath}
\usepackage{amsthm}
\usepackage{xcolor}
\usepackage{soul} 
\usepackage{balance} 
\usepackage{lastpage}  
\usepackage{verbatim}
\usepackage{mathtools}
\usepackage{ragged2e} 
\usepackage{etoolbox}
\usepackage{multirow}
\usepackage{titlesec} 	
\usepackage{changepage}
\usepackage[utf8]{inputenc}
\definecolor{linkcolor}{RGB}{7,94,84}  
\usepackage[	colorlinks=true,    	
			pdfstartview=FitV,
			linkcolor= linkcolor,
			citecolor= linkcolor,
			urlcolor= linkcolor,
			linktoc=page,
			hyperindex=true,
			hyperfigures=true,  
			breaklinks=true]
			{hyperref}
\usepackage{tikz}
\usetikzlibrary{quotes,angles}

\newcommand{\beq}{\begin{equation}}
\newcommand{\eeq}{\end{equation}}
\newcommand{\ba}{\begin{eqnarray}}
\newcommand{\ea}{\end{eqnarray}}

\definecolor{myred}{RGB}{150,50,60}

\def\red#1{{\color{black}#1}}	

\theoremstyle{remark}
\newtheorem{assumption}{Assumption}

\begin{document}

\title{\red{Entanglement and separability in continuum Rokhsar-Kivelson states}}

\author{Christian Boudreault}
\email{christian.boudreault@cmrsj-rmcsj.ca}
\affiliation{D\'epartement des sciences de la nature, Coll\`ege militaire royal de Saint-Jean,
15 Jacques-Cartier Nord, Saint-Jean-sur-Richelieu, QC, Canada, J3B 8R8}
\affiliation{Universit\'{e} de Montr\'{e}al, C.\,P.\,6128, Succursale Centre-ville, Montr\'eal, QC, Canada, H3C 3J7}

\author{Cl\'ement Berthiere}
\email{clement.berthiere@umontreal.ca}
\affiliation{Universit\'{e} de Montr\'{e}al, C.\,P.\,6128, Succursale Centre-ville, Montr\'eal, QC, Canada, H3C 3J7}
\affiliation{Centre de Recherches Math\'{e}matiques, Universit\'{e} de Montr\'{e}al, Montr\'{e}al, QC, Canada, H3C 3J7}

\author{William Witczak-Krempa}
\email{w.witczak-krempa@umontreal.ca}
\affiliation{Universit\'{e} de Montr\'{e}al, C.\,P.\,6128, Succursale Centre-ville, Montr\'eal, QC, Canada, H3C 3J7}
\affiliation{Centre de Recherches Math\'{e}matiques, Universit\'{e} de Montr\'{e}al, Montr\'{e}al, QC, Canada, H3C 3J7}
\affiliation{Institut Courtois, Universit\'e de Montr\'eal, Montr\'eal (Qu\'ebec), H2V 0B3, Canada}

\date{\today} 

\begin{abstract}
\vspace{-1pt}
\begin{center}\textbf{\abstractname}\end{center}\vspace{-5pt} 
\noindent 
\red{We study a vast family of continuum Rokhsar-Kivelson (RK) states, which have their groundstate encoded by a local quantum field theory. These describe certain quantum magnets, and are also important in quantum information. We prove the separability of the reduced density matrix of two disconnected subsystems, implying the absence of entanglement between the two subsystems---a stronger statement than the vanishing of logarithmic negativity. As a particular instance, we investigate the case where the groundstate is described by a relativistic boson, which is relevant for certain magnets or Lifshitz critical points with dynamical exponent $z=2$, and we propose nontrivial deformations that preserve their RK structure.}
Specializing to 1D systems, we study a deformation that maps the groundstate to the quantum harmonic oscillator, leading to a gap for the boson. We study the resulting correlation functions, and find that cluster decomposition is restored.
We analytically compute the $c$-function for the entanglement entropy 
along a renormalization group flow for the wavefunction, which is found to be strictly decreasing as in CFTs.
\red{Finally, we comment on the relations to certain stoquastic quantum spin chains. We show that the Motzkin and Fredkin chains possess unusual entanglement properties not properly captured by previous studies.
}

\end{abstract}

\maketitle
\makeatletter
\def\l@subsubsection#1#2{}
\makeatother
\tableofcontents

\addtolength{\skip\footins}{-10pt}

\section{Introduction}

Rokhsar-Kivelson (RK) states \cite{PhysRevLett.61.2376,henley2004classical} are groundstates that quantum mechanically encode the partition function of a classical system. They describe numerous quantum magnets at low temperature \cite{2011PhRvL.107b0402Z,2011PhRvB..84s5128S}, quantum critical phase transitions \cite{ardonne2004topological}, and novel spin chains \cite{2015PhRvB..91o5150S,2020PhRvB.101q4440H}. 
One can in principle construct a RK wavefunction for a quantum theory in $(d+1)$ spacetime dimensions from an arbitrary $d$-dimensional classical model (e.g.\;statistical mechanics). This quantum-classical correspondence can actually be made more precise, e.g., for lattice systems~\cite{henley2004classical,2005AnPhy.318..316C,2015PhRvB..91o5150S}.
The normalization factor of the (unnormalized) RK groundstate $|\Psi_0^{(d+1)}\rangle$ is interpreted as the partition function of the lower-dimensional classical system,
$\langle\Psi_0^{(d+1)}|\Psi_0^{(d+1)}\rangle=\mathcal{Z}_{\text{cl}}^{(d)}$.
Because of their simple form and relation to lower-dimensional classical theories, RK states offer a useful and controlled framework to study, e.g., correlation functions and entanglement measures such as entanglement entropy, which is notoriously hard to achieve in quantum many-body systems.
Systems poised at a RK point have been extensively studied, in particular their entanglement properties \cite{fradkin2006entanglement,PhysRevB.75.214407,PhysRevB.80.184421,Hsu:2008af,PhysRevB.82.125455,Hsu:2010ag,Oshikawa:2010kv,2011PhRvB..84s5128S,2011PhRvL.107b0402Z,St_phan_2013,Chen:2016kjp,Zhou:2016ykv,chen2017quantum,chen2017gapless,MohammadiMozaffar:2017nri,Angel-Ramelli:2019nji,angel-ramelli2020logarithmic,Angel-Ramelli:2020xvd}.

In this work, we study the entanglement and correlations properties of a vast family of {\slshape continuum} RK states for which the dual classical models are local quantum field theories (QFTs).
We prove the separability of the reduced density matrix of two disconnected subsystems for continuum RK states, implying the absence of entanglement between the two subsystems---a stronger statement than the vanishing of logarithmic negativity \cite{HORODECKI19961,Peres:1996dw,Vidal:2002zz,Plenio:2005cwa}.

As a particular instance, we investigate Lifshitz groundstates with dynamical exponent $z=2$. We propose nontrivial deformations that preserve their RK structure, making these theories rare examples of nonrelativistic theories which admit analytic treatment. 
By examining critical Lifshitz theories, comparing entanglement entropy and capacity of entanglement we argue that entanglement is effectively carried by maximally entangled EPR pairs, unlike systems with emergent Lorentz invariance (e.g. gapless Dirac fermions).

We also apply our findings to certain stoquastic spin chains introduced by Shor, Bravyi et al \cite{Bravyi2012criticality}. We show that Motzkin and Fredkin spin chains possess unusual entanglement properties not properly captured by continuum descriptions proposed in previous studies. Our work raises the urgent question regarding the correct field theoretical description of those spin chains, and we conjecture possibilities.

This paper is organized as follows. We begin in Section \ref{S:RK} with some background on (continuum) RK states, and study their reduced density matrices. We then prove the separability of the reduced density matrix of two disconnected subsystems for continuum RK wavefunctional. In Section \ref{S:QLT}, we review key concepts for Lifshitz theories, and in particular the RK structure of their groundstates. We introduce in Section \ref{S:mass} a massive deformation preserving the RK property of the Lifshitz groundstate, and then proceed to study this model in $1+1$ dimensions. We present our results on correlations and entanglement properties of the (massive) theory. We find that cluster decomposition, violated in the massless case, is restored by the regulating mass. 
Along the way, aside from Rényi entropies and mutual information, we compute other entanglement-related quantities: the capacity of entanglement, the entropic $c$-function, and the sharp limit coefficient of corner entanglement. Our results on the positive-field version of the massless Lifshitz theory are presented in Section \ref{S:positive}, where we discuss its relation to Motzkin and Fredkin spin chains and results in the literature. We conclude in Section \ref{S:conclu} with a summary of our main results, and give an outlook on future study. Appendix \ref{S:singular} is devoted to a non-Gaussian deformation of Lifshitz theory for which the groundstate is described by $SL(2,\mathbb{R})$-conformal quantum mechanics. This deformation preserves the conformal spatial symmetry of the groundstate, and constrains the form of the correlators and Rényi entropies. Finally, Appendix \ref{A:capacity} contains our results on the capacity of entanglement.

\section{Rokhsar-Kivelson states, reduced density matrices, and separability}
\label{S:RK}

In this work, we study the vast family of quantum RK states, that is, groundstates of the form 
\beq\label{RK}
\Psi_0[\phi] \propto e^{-\frac{1}2S_{\text{cl}}[\phi]}\,,
\eeq
for some \textit{local} Euclidean action $S_{\text{cl}}[\phi]$ \red{involving a finite number of derivatives of the field}. Such a state is the groundstate of a parent Hamiltonian
\beq
H=\frac{1}{2}\int d^d x \;\{A^{\dagger}(x),A(x)\}\,,
\eeq
quadratic in the operators \beq
A(x)=\frac{1}{\sqrt{2}}\left(i\Pi (x)+\frac{1}2\frac{\delta S_{\text{cl}}[\phi]}{\delta\phi(x)}\right),
\eeq
where, in the field-eigenstate Schrödinger picture, canonical quantization demands $\Pi(x)=-i\delta / \delta\phi(x)$. Clearly, positive semidefiniteness is preserved for any deformation of the type
\beq
\frac{1}{2}S_{\text{cl}} \mapsto \frac{1}{2}S_{\text{cl}} +\Lambda \,,
\eeq
where the deformations $\Lambda$ is real and local to ensure reality and locality of the action. With $A(x)$ mapped to
\beq\label{E:transfoA}
A'(x) = A(x)+\tfrac{\delta}{\delta\phi(x)}\Lambda = e^{-\Lambda}A(x)e^{\Lambda}\,,
\eeq
we see that (normalizable) states annihilated either by the $A(x)$'s or the $A'(x)$'s are related under the transformation $\Psi_0' = e^{-\Lambda}\Psi_0$:
\beq
A(x)\Psi_0[\phi] = 0 \Longleftrightarrow A'(x)\Psi'_0[\phi] = 0 \,.
\eeq
Such states, if any, must be groundstates of their respective Hamiltonians.
We emphasize that the new Hamiltonian $H'$ is generally \textit{not} the result of a simple frame transformation, as $H'\neq e^{-\Lambda}H e^{\Lambda}$. ($H'$ and $e^{-\Lambda}He^{\Lambda}$ agree if and only if $\Lambda$ is imaginary-valued.) 

In this work, we shall study a particular instance of RK states, that is Lifshitz groundstates and perform deformations thereof.
In the following section, we first derive general results for continuum RK wavefunctional, such as a formula for Rényi entanglement entropies, and show the separability of the reduced density matrix of disconnected subsystems.

\subsection{Reduced density matrix and replica trick}\label{A:replica_trick}

We derive formulas for the Rényi entanglement entropies of continuum RK wavefunctional. For simplicity, we work in one spatial dimension, though our results generalize naturally to higher dimensions.
The Rényi entanglement entropies for a finite bipartition $\{A,B\}$ are defined as
\beq\label{E:renyi_def}
\begin{aligned}
S_n (A)&= \frac{1}{1-n}\log \text{Tr}\,\rho^n_A\,,
\end{aligned}
\eeq
where $\rho_A$ is the reduced density matrix on subsystem $A$. The entanglement entropy is obtained from \eqref{E:renyi_def} by taking the replica limit $n\rightarrow 1$.

We begin with normalized states formally written as $|\Psi\rangle=\int\mathcal{D}\phi \; \Psi[\phi]\,|\phi\rangle$, in one-to-one correspondence with Schrödinger wavefunctionals $\Psi$, and we make two assumptions:
\begin{assumption}
The $\phi$ are (dimensionless) real fields, almost everywhere $C^1$ on $[0,L]$, and $\mathcal{D}\phi$ is a measure for such fields.
\end{assumption}
Field eigenstates
\beq
\chi_{\phi}[\phi']=\begin{cases}
	1\;, & \quad\phi'=\phi,\\
	0\;, & \quad\text{otherwise},
\end{cases}
\eeq
have no spatial entanglement: for almost any open partition $\{A,B\}$ of the physical space, $\chi_{\phi}=\chi_{\phi |_A}\otimes \chi_{\phi |_B}$. Equivalently,
\beq\label{E:appendix_alpha_beta}
|\phi\rangle  = |\alpha\rangle_{\pmb{\phi}}\otimes|\beta\rangle_{\pmb{\phi}}\,,
\eeq
where $\alpha=\phi|_A$ and $\beta=\phi|_B$ are $C^1$ almost everywhere on their respective domain, and $\pmb{\phi}=\{ \phi_1 , \dots , \phi_M \}$ is an \textit{anchor} for these fields: $\alpha(x_i)=\phi_i = \beta(x_i)$ for each $x_i$ at the common boundary of $A$ and $B$. We now make an assumption specific to a distinguished  \red{class of states} $|\Psi\rangle$:
\begin{assumption}
$\Psi$ is real-valued, and for $\alpha , \beta$ as above, $\Psi[\phi]=a[\alpha]\otimes b[\beta]$ for some real functionals $a,b$. \red{This means that the state $\Psi$ is \emph{local}. } 
\end{assumption}
Then 
\beq\label{E:Schmidt_formal}
\begin{aligned}
|\Psi\rangle &=\int d\pmb{\phi}\int (\mathcal{D}\alpha)_{\pmb{\phi}}\; a[\alpha]|\alpha\rangle_{\pmb{\phi}} \otimes \int (\mathcal{D}\beta)_{\pmb{\phi}}\; b[\beta]|\beta\rangle_{\pmb{\phi}}\\
		&=\int d\pmb{\phi}\;|\Psi_A\rangle_{\pmb{\phi}}\otimes|\Psi_B\rangle_{\pmb{\phi}}\,,
\end{aligned}
\eeq
which formally resembles a Schmidt decomposition for $|\Psi\rangle$ with anchors as Schmidt index, and can be cast in actual Schmidt form $|\Psi\rangle=\sum_{i=1}^{\infty} \sigma_i |u_i\rangle \otimes |v_i\rangle$, if necessary, by expressing the $\pmb{\phi}$-integral as a convergent sum over increasingly finer mesh elements such that the $i$th mesh element contains point $\pmb{\phi}_i$, and where $\{|u_i\rangle\}$ and $\{|v_i\rangle\}$ are respectively obtained by Gram-Schmidt orthonormalization of the sets
$\{|\Psi_A\rangle_{\pmb{\phi}_i}-\sum_{j\prec i} |\Psi_A\rangle_{\pmb{\phi}_j}\}$, and $\{|\Psi_B\rangle_{\pmb{\phi}_i}-\sum_{j\prec i} |\Psi_B\rangle_{\pmb{\phi}_j}\}$. (Here, $j\prec i$ means that $j$ is a coarser mesh element containing $i$.) We will not require the Schmidt decomposition, and will be content with~\eqref{E:Schmidt_formal}. Actually, we will use Assumption 2 only at the end of the argument, so for now we simply write the state in terms of anchored fields, $|\Psi\rangle=\int d\pmb{\phi}\int (\mathcal{D}\alpha\beta)_{\pmb{\phi}}\; \Psi[\alpha\beta]|\alpha\rangle_{\pmb{\phi}} \otimes |\beta\rangle_{\pmb{\phi}}=\int d\pmb{\phi} \; |\Psi\rangle_{\pmb{\phi}}$. 
\red{Note that in higher dimensions, integrals over fields become path-integrals.}
The corresponding density matrix is
\beq
\begin{aligned}
\rho&=|\Psi\rangle\langle\Psi|\\
	&=\int (d\pmb{\phi}\pmb{\phi'})(\mathcal{D}\alpha\beta)_{\pmb{\phi}}(\mathcal{D}\alpha'\beta')_{\pmb{\phi'}} \Psi[\alpha\beta] \Psi^* [\alpha'\beta']\\
	&\hspace{80pt}\times |\alpha\rangle_{\pmb{\phi}} |\beta\rangle_{\pmb{\phi}}\,{}_{\pmb{\phi'}}\langle\alpha'|{}_{\pmb{\phi'}}\langle\beta'|\,,
\end{aligned}
\eeq
and the reduced density is
\beq\label{E:reduced_density_appendix}
\begin{aligned}
\rho_A&=\text{Tr}_B \,\rho\\
	&=\int d\pmb{\phi}(\mathcal{D}\alpha\alpha' \beta)_{\pmb{\phi}}\; |\alpha\rangle_{\pmb{\phi}} \; \big(\Psi[\alpha\beta]\Psi^* [\alpha' \beta]\big)\;{}_{\pmb{\phi}}\langle\alpha'|\,.
\end{aligned}
\eeq
Note that all fields now have the same anchor due to the partial trace $\text{Tr}_B (\cdot )=\int d\pmb{\phi} (\mathcal{D}\beta)_{\pmb{\phi}}\;{}_{\pmb{\phi}}\langle\beta|\cdot |\beta\rangle_{\pmb{\phi}}$. When $\text{Tr}_A \, \rho_A^n$ is explicitly written out, we find a cyclic product of projectors of the form
\beq
\begin{aligned}
\text{Tr}_A &\left(\cdots |\alpha_i\rangle_{\pmb{\phi}_i}{}_{\pmb{\phi}_i}\langle\alpha'_i |\alpha_{i+1}\rangle_{\pmb{\phi}_{i+1}}{}_{\pmb{\phi}_{i+1}}\langle\alpha'_{i+1} |\cdots\right)\\
	&\quad=\prod_i \delta_{\alpha'_{i}\alpha_{i+1}}\delta_{\pmb{\phi}_i \pmb{\phi}_{i+1}}\,,
\end{aligned}
\eeq 
forcing all anchors to agree: $\pmb{\phi}_1=\cdots =\pmb{\phi}_n=\pmb{\phi}$. Thus
\beq
\begin{aligned}
\text{Tr}_A \, \rho_A^n &= \int d\pmb{\phi} \prod_i \int (\mathcal{D}\alpha_i\beta_i)_{\pmb{\phi}}\; |\Psi[\alpha_i \beta_i]  |^2\\
	&= \int d\pmb{\phi}\left[\int (\mathcal{D}\alpha\beta)_{\pmb{\phi}}\; |\Psi [\alpha\beta]|^2\right]^n\\
	&=\int d\pmb{\phi}\;\Big[{}_{\pmb{\phi}}\langle\Psi |\Psi\rangle_{\pmb{\phi}}\Big]^n.
\end{aligned}
\eeq
Manifestly, $\text{Tr}_A \, \rho_A^n = \text{Tr}_B \, \rho_B^n$. Up to this point we have only used Assumption 1. If we now use the second assumption, we obtain
\beq\label{E:app_AB}
\text{Tr}_A \, \rho_A^n = \int d\pmb{\phi}\left[\int (\mathcal{D}\alpha)_{\pmb{\phi}}\; a^2[\alpha] \int (\mathcal{D}\beta)_{\pmb{\phi}}\; b^2[\beta]\right]^n.
\eeq
Furthermore, on any interval $U=(x_{i_1} , x_{i_2})$ for $x_i$ at the common boundary of $A$ and $B$, the functional $\Psi$ factors as $\Psi |_U \times \cdots$, and we may consistently define a local ``action" by adjoining pieces of the form
\beq
\frac{1}{2} S_U =-\log \Psi |_U \,.
\eeq
Then \eqref{E:app_AB} splits into factors $\langle \phi_{i_2} , x_{i_2} | \phi_{i_1} , x_{i_1}\rangle \overset{\text{def}}{=} \int (\mathcal{D}\alpha)_{\pmb{\phi}}\; (\Psi |_U)^2[\alpha]=\int (\mathcal{D}\alpha)_{\pmb{\phi}}\; e^{-S_U [\alpha]}$, which we identify as the propagators of the dynamics associated to $\Psi$. It follows that
\beq\label{E:Tr_versus_fM}
\text{Tr}_A \, \rho_A^n = \int d\pmb{\phi}\left[ f_M (\pmb{\phi} , \mathbf{x})\right]^n,
\eeq
with $f_M (\pmb{\phi} , \mathbf{x})=\langle h_2 , L | \phi_M , x_M\rangle \cdots \langle \phi_1 , x_1 | h_1 , 0\rangle$ for boundary conditions $\phi(0)=h_1$, $\phi(L)=h_2$. If the system is periodic, the (variable) boundary value $\phi_0=\phi(0)=\phi(L)$ is an additional anchor to be integrated on in \eqref{E:Tr_versus_fM}.

For the dimensionful fields, as in~\eqref{E:dim_phi}, we have $\text{dim }f_M=(\text{length})^{-M/2}$. Under the rescaling $\phi\to\sqrt{\epsilon}\phi$, $f_M\to\epsilon^{-M/2}f_M$, where $\epsilon$ is a local length scale (e.g. a lattice constant or UV cutoff), we have
\beq
\int d\pmb{\phi} \;[f_M]^n \to \epsilon^{(n-1)M/2}\int d\pmb{\phi}\; [f_M]^n,
\eeq
which, using \eqref{E:Tr_versus_fM} in \eqref{E:renyi_def}, yields the Rényi entropies
\beq
S_n (A)=\frac{1}{1-n}\log\Big[\epsilon^{(n-1)M/2}\int d\pmb{\phi}\; \left[f_M(\pmb{\phi}, \mathbf{x})\right]^n\Big]\,.
\eeq

\red{To eliminate any concerns about normalization, one can construct the entropy out of the explicitly normalized object $\rho/\text{Tr}\,\rho$, which amounts to making the change $f_M(\pmb{\phi}, \mathbf{x})\rightarrow f_M(\pmb{\phi}, \mathbf{x})/\int d\pmb{\phi}'f_M(\pmb{\phi}', \mathbf{x})$ in the above formula.}

\subsection{Separability of $\rho_{A\cup B}$ for disjoint subsystems}\label{A:separability}
We show here that for general continuum states satisfying the RK property (which include the models considered next in the present work), the reduced density $\rho_{A\cup B}$ is (mixed) separable,
\beq\label{E:mixed_separable}
\rho_{A\cup B}=\sum_{j\in J} p_j (\rho_{A,j}\otimes \rho_{B,j})\,,
\eeq
for any disconnected subsystems $A$ and $B$. In the above expression, $J$ is some index set, and $p$ is a probability distribution on $J$. Thus, tracing out the complement of $A \cup B$ completely disentangles $|\Psi_0\rangle$. However, as long as $p$ is not the trivial distribution, $A$ and $B$ will share mutual information.
 
Consider a general tripartition $\{A,C,B\}$, where $C$ is to be traced out. From \eqref{E:reduced_density_appendix}, a state satisfying both assumptions of Section~\ref{A:replica_trick} has reduced density
\beq
\langle \alpha\beta | \rho_{A\cup B}|\alpha' \beta' \rangle = \int (\mathcal{D}\gamma)_{\pmb{\phi}} \Psi[\alpha\beta\gamma]\Psi[\alpha' \beta' \gamma] \,,
\eeq 
where $\alpha, \beta$ are as in~\eqref{E:appendix_alpha_beta}, $\pmb{\phi}$ is the anchor for these fields, and $\gamma$ is restricted to the complement of $A\cup B$. By the second assumption, $\Psi[\alpha\beta\gamma]=a[\alpha]b[\beta]\Gamma[\gamma]$, where $\Gamma$ is the restriction of  $\Psi$ to the complement of $A\cup B$, so
\beq
\langle \alpha\beta | \rho_{A\cup B}|\alpha' \beta' \rangle = \int (\mathcal{D}\gamma)_{\pmb{\phi}} a[\alpha] b[\beta] \Gamma^2 [\gamma]a[\alpha'] b[\beta'] \,.
\eeq 
\red{Since $A$ and $B$ do not have a boundary in common, they do not share any anchor. Therefore, the reduced density is a separable mixed state as it can be written as}
\beq\label{E:separable_appendix}
\rho_{A\cup B} = \int d\pmb\phi \hspace{3pt} p(\pmb\phi) \hspace{3pt} \rho_A (\pmb\phi)\otimes \rho_B (\pmb\phi)\,,
\eeq
with probability distribution
\beq\label{E:pdf_separability}
p(\pmb\phi) =\int (\mathcal{D} \alpha\beta\gamma)_{\pmb\phi} \; a^2[\alpha]b^2[\beta]\Gamma^2 [\gamma] \,,
\eeq
and subsystem densities
\beq\label{E:sub_densities_appendix}
\begin{aligned}
\rho_A (\pmb\phi) &= \frac{\int (\mathcal{D} \alpha \alpha')_{\pmb\phi}\; |\alpha\rangle_{\pmb\phi} a[\alpha]a[\alpha'] {}_{\pmb{\phi}}\langle\alpha' |}{\int (\mathcal{D} \alpha)_{\pmb\phi} \; a^2[\alpha]}=        |\Psi_A\rangle_{\pmb{\phi}}{}_{\pmb{\phi}}\langle\Psi_A |\,,\\
\rho_B (\pmb\phi) &= \frac{\int (\mathcal{D} \beta \beta')_{\pmb\phi} \;|\beta\rangle_{\pmb\phi} b[\beta]b[\beta'] {}_{\pmb{\phi}}\langle\beta' |}{\int (\mathcal{D} \beta)_{\pmb\phi} \; b^2[\beta]}=
|\Psi_B\rangle_{\pmb{\phi}}{}_{\pmb{\phi}}\langle\Psi_B |\,.
\end{aligned}
\eeq
\red{Note that \eqref{E:pdf_separability} is a probability distribution by virtue of the normalization of $|\Psi\rangle$, and that the vectors $|\Psi_A\rangle_{\pmb{\phi}}$ and $|\Psi_B\rangle_{\pmb{\phi}}$ in \eqref{E:sub_densities_appendix} are normalized by construction.} If necessary, the integral in~\eqref{E:separable_appendix} can be converted to a convergent countable sum over increasingly fine mesh elements, in the same way the (discrete) Schmidt form was obtained in Section~\ref{A:replica_trick}. 
As obvious from the subsystem densities~\eqref{E:sub_densities_appendix}, $\rho_{A\cup B}$ is invariant under partial transpositions
\beq\label{E:PT_appendix}
\rho_{A\cup B}^{T_A}=\rho_{A\cup B}=\rho_{A\cup B}^{T_B}\,.
\eeq
This is consistent with the known fact that, for Gaussian states like the massive deformation groundstate considered in the main text, invariance under partial transposition implies separability \cite{lami2018gaussian}. That observation was used in \cite{angel-ramelli2020logarithmic} to prove mixed separability in the massless Lifshitz theory. Our treatment is quite general, valid for real-valued RK wavefunctionals, independently of the dimension of the underlying classical theory. It thus provides the explicit separability of $\rho_{A\cup B}$ for quantum states satisfying the aforementioned properties, which includes both the massive deformation (Gaussian) and the singular deformation (non-Gaussian) of the Lifshitz theory considered in this work. \red{Separability implies the vanishing of logarithmic negativity.}

\red{Our result on the separability of continuum RK states for disconnected subsystems $A$ and $B$ should carry through to the case where the fields are compact, i.e.\;for $\phi\sim\phi + 2\pi R$, with $R>0$ being the compactification radius. Indeed, since $A$ and $B$ do not have a common boundary, they do not share any anchor so that the compacity of the field, which manifests itself in the boundary conditions, should not affect our argument to obtain a separable reduced density \eqref{E:separable_appendix}. We emphasize that the crucial point is the wavefunctional being a local function of the field such that the fields on $A,B,C$ can only talk through the boundaries. It can also be shown that the logarithmic negativity vanishes in the compact case. Consider its replica formulation \cite{Calabrese:2012ew}, $\mathcal{E}(A:B)=\lim_{n_e \to 1}\log {\rm Tr}\big(\rho_{A\cup B}^{T_A} \big)^{n_e}$, where the analytic continuation is taken over the even integers $n_e$. Taking $n_e$ copies of $\rho_{A\cup B}^{T_A}$, the sewing conditions between the copies make all anchors agree, i.e.\;all the replica fields agree at the boundary between $C$ and $A,B$. Furthermore, the partial transposition $\big(\rho_{A\cup B}^{T_A} \big)^{n}$ is not sensitive to the parity of $n$, and one can show that ${\rm Tr}\big(\rho_{A\cup B}^{T_A} \big)^{n}={\rm Tr} \big(\rho_{A\cup B} \big)^{n}$. 
Taking the limit $n\rightarrow 1$, it then follows from the unit normalization of the density matrix (${\rm Tr}\,\rho_{A\cup B}^{T_A}={\rm Tr}\,\rho_{A\cup B}=1$) that the topological (winding) sector contribution coming from the compact nature of the field is trivial, yielding a zero logarithmic negativity (see \cite{angel-ramelli2020logarithmic} for an explicit calculation for Gaussian groundstates of certain Lifshitz theories in one and two spatial dimensions).}

\section{Lifshitz critical point}\label{S:QLT}

\red{Scale invariance plays a central role in the study of dynamical critical phenomena, far-from-equilibrium statistical dynamics, and quantum criticality \cite{hohenberg1977theory,cardy1996scaling,marro1999nonequilibrium,sachdev2011quantum}. Taken isotropic, scale invariance is often enhanced to conformal symmetry, with profound consequences. However, many systems at criticality exhibit anisotropic scaling between space and time, called Lifshitz scaling,
\beq\label{E:lifshitz_scaling}
t\to\lambda^z t \; , \quad\; \mathbf{x}\to\lambda\mathbf{x}\; , \quad\;\;  \lambda>0 \,,
\eeq 
with characteristic dynamical critical exponent $z\neq1$. Lifshitz scaling is encountered in a variety of contexts, from nonrelativistic mechanics \cite{dealfaro1976conformal,hagen1972scale,jackiw1972introducing,romero2011conformal} and critical systems \cite{henkel2002phenomenology,ardonne2004topological}, to nonrelativistic holographic duality \cite{son2008toward,balasubramanian2008gravity,barbon2008on,bertoldi2008thermodynamics,keranen2017correlation} and  quantum gravity \cite{horava2009quantum,horava2009membranes}. Here, we are interested in a certain class of nonrelativistic quantum field theories admitting Lifshitz symmetry. Originally introduced as the `quantum Lifshitz model' in $2+1$ dimensions for $z=2$ \cite{ardonne2004topological}, general $(d+1)$-dimensional Lifshitz theories with (even) positive integer $z$ possess the remarkable feature that their groundstate wavefunctional takes a local form, given in terms of the action of a $d$-dimensional classical model---a RK wavefunctional.}

The real, noncompact $(d+1)$-dimensional $z=2$ Lifshitz quantum critical boson is the QFT with Hamiltonian
\beq\label{E:H_Lifshitz}
H=\frac{1}{2}\int d^d x \left(\Pi^2 + \kappa^2(\nabla^2 \phi)^2 \right),
\eeq
with canonical commutation relations $[\phi(x),\Pi(x')]=i\delta(x-x')$. The parameter $\kappa$ is dimensionless, and $\Pi(x)=-i\delta / \delta\phi(x)$ in the Schrödinger picture. In addition to invariance under Lifshitz scaling~\eqref{E:lifshitz_scaling}, and the obvious $\mathbb{Z}_2$ symmetry, this theory is invariant under affine shifts of the field $\phi(x)\to\phi(x)+ax+b$. This is also called polynomial shift symmetry.

The Hamiltonian~\eqref{E:H_Lifshitz} is quadratic in the operators
\beq\label{E:operator_A}
A(x)=\frac{1}{\sqrt{2}}\left(i\Pi (x)-\kappa \nabla^2\phi\right),
\eeq
as it is easily verified that $H=\frac{1}{2}\int d^d x \;\{A^{\dagger}(x),A(x)\}$.
Note that because $\Pi(x)$ is Hermitian, $A^{\dagger}(x)$ is obtained from~\eqref{E:operator_A} by the single replacement $i\to -i$. Alternatively, it will be convenient to diagonalize the Hamiltonian to the normal-ordered form 
\beq\label{E:H_normal_Lifshitz}
H= \int d^d x \;A^{\dagger}(x)A(x)+ E_{\text{vac}}\,,
\eeq
where $E_{\text{vac}}=\frac{1}{2}\int d^d x [A(x),A^{\dagger}(x)]=-\frac{\kappa}{2}\int d^d x \,\nabla_x^2 \delta^d (x-y)|_{x=y}$ is a positive, UV-divergent multiple of the identity. As such, it can be considered as a vacuum-energy shift relating the otherwise identical eigensystems of $H$ and $\int d^d x \,A^{\dagger}(x)A(x)$ \cite{ardonne2004topological}. A groundstate is now found satisfying
\beq
A(x)|\Psi_0 \rangle =0\; , \quad \forall x\,.
\eeq
Positive semidefiniteness of $\int d^d x \,A^{\dagger}(x)A(x)$ ensures that such a state is indeed a groundstate. The corresponding functional-differential equation, $(\frac{\delta}{\delta \phi}-\kappa\nabla^2 \phi)\Psi_0[\phi]=0$, has nontrivial solution
\beq\label{E:Psi_0}
\Psi_0[\phi]=\frac{1}{\sqrt{\mathcal{Z}}}\hspace{1pt}e^{-\frac{1}{2}\int d^d x \; \kappa (\nabla\phi)^2}\,,
\eeq
with normalization factor
\beq\label{E:Z}
\mathcal{Z}=\int \mathcal{D}\phi(x)\; e^{-\int d^d x \;\kappa (\nabla\phi)^2}.
\eeq
One recognizes $\mathcal{Z}$ as the partition function of a $d$-dimensional free Euclidean scalar field with classical action $S_{\text{cl}}[\phi]=\int d^d x \,\kappa (\nabla\phi)^2$. This local action appearing in $\Psi_0$ is conformally invariant in $d$ (spatial) dimensions. We thus have an emergent \textit{spatial} conformal symmetry in the groundstate of the parent Hamiltonian $H$. We emphasize that $S_{\text{cl}}$ is only inherent to $\Psi_0$, and does not coincide with the action of the parent Hamiltonian $H$. (A relationship between these actions can be established via stochastic quantization \cite{dijkgraaf2010relating}.) Dimensionlessness of $S_{\text{cl}}$ requires that
\beq\label{E:dim_phi}
\text{dim}\, \phi = (\text{length})^{-\frac{d}{2}+1},
\eeq
and Lifshitz scaling in turn implies
\beq\label{E:dim_Pi}
\text{dim}\, \Pi = (\text{length})^{-\frac{d}{2}-1}.
\eeq
Equivalently, the dimension of $H$ is $(\text{length})^{-2}$ as expected from the Lifshitz scaling $t\sim x^z$. For simplicity, we will mostly consider the case $d=1$, for which $S_{\text{cl}}$ is the Euclidean action of a free nonrelativistic particle. We will derive certain properties of the entanglement entropy of the $d=2$ theory that follow from our $d=1$ results in Section~\ref{S:2+1_cornerEE}. Using the Gaussian propagators of this underlying theory,
\beq\label{E:propagator_Lif}
\langle \phi' , x' | \phi , x \rangle = \sqrt{\frac{\kappa}{\pi (x'-x)}}e^{-\frac{\kappa (\phi'-\phi)^2}{x'-x}},
\eeq 
with $x' >x$, one can compute groundstate correlation functions, and Rényi entanglement entropies \cite{chen2017gapless}. 

In this work, we perform two distinct deformations of the Lifshitz point preserving the Rokhsar–Kivelson structure. In Section \ref{S:mass} we consider a deformation that breaks the emergent conformal symmetry of $S_{\text{cl}}[\phi]$ with a (mass) scale, whereas in Appendix \ref{S:singular} we consider a nontrivial deformation which preserves this symmetry when $d=1$.  (The affine field-shift symmetry is lost in both cases, and will play no further role.)

\section{Deformation by a mass term}\label{S:mass}
We now explicitly break the Lifshitz scaling symmetry of~\eqref{E:H_Lifshitz} with a length scale $1/m$, setting $\Lambda[\phi]=\frac{1}{2}\int d^dx \, m^2\phi(x)^2$, so that the Euclidean action defining the groundstate becomes 
\beq\label{E:S_mass_KG}
S_{\text{cl}}[\phi]=\int d^dx \left(\kappa (\nabla\phi)^2 + m^2\phi^2\right),
\eeq
and
\beq
A(x) =\frac{1}{\sqrt{2}}\left(\frac{\delta}{\delta\phi(x)}-\kappa \nabla^2\phi +m^2\phi\right).
\eeq
Expression \eqref{E:S_mass_KG} is seen to correspond to the Euclidean action of a massive relativistic scalar (Klein-Gordon). The original Lifshitz Hamiltonian~\eqref{E:H_Lifshitz} is thereby deformed to
\beq\label{E:H_Lifshitz_mass}
H_{\text{mass}}=H_{\text{Lif}}+\frac{1}{2}\int d^d x \left(2\kappa m^2 (\nabla \phi)^2 + m^4 \phi^2\right).
\eeq
The Hamiltonians $H_{\text{mass}}=\frac{1}{2}\int d^d x \;\{A^{\dagger}(x),A(x)\}$ and $H_{\text{mass}}^{\text{normal}}=\int d^d x \;A^{\dagger}(x)A(x)$ differ by $\frac{1}{2}\int d^d x [A^{\dagger}(x),A(x)]=\frac{1}{2}\text{tr}(\kappa \nabla^2 - m^2)$, a UV-divergent multiple of the identity. Their eigensystems are thus identical up to an infinite zero-point energy shift, with common groundstate
\beq\label{E:A_Psi_mass}
\Psi_0[\phi] = \frac{1}{\sqrt{\mathcal{Z}}}\hspace{1pt}e^{-\frac{1}{2}\int d^d x \left(\kappa (\nabla\phi)^2+m^2\phi^2\right)}\, .
\eeq
Indeed it is clear that $A(x)|\Psi_0 \rangle =0$. Due to the scale $m$, the new groundstate has lost conformal symmetry. $H_{\text{mass}}$ has positive-infinite groundstate energy $E_{\text{vac}}=\frac{1}{2}\text{tr}(-\kappa \nabla^2 + m^2)$, whereas the normal-ordered form has groundstate energy zero. The new terms in~\eqref{E:H_Lifshitz_mass} are both relevant under renormalization group (RG), and the theory flows to the massive relativistic scalar in the IR. Let us be more precise: the usual RG transformation on the theory sends $x_i\to b x_i$ ($1\leq i\leq d$) and $t\to b^z t$, with $z=2$. The scaling dimensions of $(\nabla\phi)^2$ and $\phi^2$ are distinct, meaning that the fine-tuning between the mass-dependent terms in \eqref{E:H_Lifshitz_mass} would be lost under RG.
The key to preserve the RK structure is to consider an alternative RG transformation that acts on the groundstate wavefunction \eqref{E:A_Psi_mass}, not directly on the Hamiltonian. This wavefunction RG transformation
now acts on the $(0+d)$D relativistic massive Euclidean scalar. All coordinates are dilated by the same factor,
$x_i\to b x_i$. The mass is relevant, and leads to the usual infinite mass trivial IR fixed point in $(0+d)$D. The wavefunction RG transformation corresponds to the trajectory in theory space where $m$ increases in \eqref{E:H_Lifshitz_mass}. In other words, the trajectory in the space of Hamiltonians is obtained by constructing the parent Hamiltonian for the groundstate \eqref{E:A_Psi_mass}. 

The normalization factor for $\Psi_0$ coincides with the Euclidean partition function of the $d$-dimensional massive scalar,
\beq\label{E:Z_m}
\mathcal{Z}=\int \mathcal{D}\phi\; e^{-\int d^d x \left(\kappa (\nabla\phi)^2+m^2\phi^2\right)}.
\eeq
The density matrix operator corresponding to $\Psi_0$ is
\beq\label{E:rho}
\rho=\frac{1}{\mathcal{Z}}\int\mathcal{D}\phi\mathcal{D}\phi' \; e^{-(S_{\text{cl}}[\phi]+S_{\text{cl}}[\phi']) /2}|\phi\rangle\langle\phi'| \, .
\eeq
We now specialize to $d=1$ with Dirichlet boundary conditions (BC): $\phi(0)=0=\phi(L)$, but many of our results will also apply to other choices of boundary conditions. 
Apart from simplicity, one motivation for this choice is the search for the elusive effective field theories of the Motzkin and Fredkin spin chains \cite{Bravyi2012criticality,DellAnna2016violation} whose unique, highly entangled, frustration-free groundstates reproduce the logarithmic scaling of entanglement entropy found in critical spin chains.  
When $\phi$ is used to represent a ``height field" for the spin variables, i.e. $\partial_x \phi(x)=S_z(x)$, the groundstate property $S_z^{\text{tot}}=0$ translates into Dirichlet boundary conditions (up to a constant) on $\phi$. The \textit{positive}-field version of~\eqref{E:H_Lifshitz} is a parent Hamiltonian for the \textit{positive}-field version of~\eqref{E:Psi_0}, which captures many spin and entanglement features of the Motzkin and Fredkin groundstates \cite{chen2017quantum, chen2017gapless, movassagh2017entanglement}, while having a markedly different excitation spectrum \cite{chen2017quantum, chen2017gapless}. 
We note that bulk properties are expected not to depend on the field positivity constraint~\cite{chen2017quantum,chen2017gapless}.

In one dimension, $\mathcal{Z}$ is the partition function of a single particle with Euclidean Lagrangian $\mathcal{L}_m=\kappa (\partial_x \phi)^2 + m^2\phi^2$, i.e. a quantum harmonic oscillator of ``mass" $M=2\kappa$ and ``frequency" 
\beq
\omega = m/\sqrt{\kappa} \,.
\eeq 
The standard propagator of the Euclidean oscillator, $\langle \phi' , x' | \phi , x \rangle=\int_{\phi(x)=\phi}^{\phi(x')=\phi'}\mathcal{D}\phi \; e^{-\int_x^{x'}dx\; \mathcal{L}_m}$,  is \cite{Ingold2002path}
\beq\label{E:propagator_mass}
\begin{aligned}
\langle \phi' , x' &| \phi , x \rangle =\sqrt{\frac{\kappa\omega}{\pi  \sinh \omega (x' - x)}}\\
&\times\exp\bigg[
					\frac{-\kappa\omega\big((\phi^2 + {\phi'}^2)\cosh \omega (x' - x) - 2\phi\phi'\big)}{\sinh\omega (x'-x)}\bigg],
\end{aligned}
\eeq 
which reduces to \eqref{E:propagator_Lif} in the limit $m=\omega=0$. Vacuum expectation values of local operators $ \hat{\mathcal{O}}_x |\phi\rangle$ can be expressed in terms of the propagator associated to $\Psi_0$ via the mapping $\mathcal{Z}\langle\Psi_0 | \Psi_0\rangle=\int_{\text{BC}} \mathcal{D}\phi\; e^{-\int d^d x S_{\text{cl}}[\phi]}$. In particular we obtain from \eqref{E:rho}
\beq
\langle\phi(x)\rangle = \text{Tr}[\rho \hat\phi(x)]=\int_{-\infty}^{\infty}d\phi \; \phi \; f_1 (\phi , x) \,,
\eeq
with single-point probability distribution $f_1 (\phi , x)=\mathcal{Z}^{-1}\langle 0,L|\phi,x\rangle\langle\phi,x|0,0\rangle$. Since $f_1$ is symmetric under $\phi\to -\phi$, field and gradient have vanishing vacuum expectation values, 
\beq
\begin{aligned}
&\langle\phi(x)\rangle  =\int_{-\infty}^{\infty}d\phi \; \phi \; f_1 (\phi , x) = 0\,,\\
&\langle\partial_x\phi(x)\rangle   = \partial_x\langle\phi(x)\rangle = 0\,.
\end{aligned}
\eeq
Other boundary conditions not breaking $\mathbb{Z}_2$ symmetry, periodic conditions for instance, will have the same vanishing expectations. 

We may formally define the reduced density matrix on the single-point set $\{x\}$ as 
\beq
\rho(x)=\int_{-\infty}^{\infty}d\phi \; f_1 (\phi , x) |\phi , x\rangle \langle\phi , x|\, .
\eeq
Then for a local operator $ \hat{\mathcal{O}}_x$ as above, $\langle\hat{\mathcal{O}}_x\rangle = \text{Tr}_{\{x\}}[ \rho (x)\hat{\mathcal{O}}_x]=\int_{-\infty}^{\infty}d\phi\;
f_1 (\phi , x)\langle \phi,x|\hat{\mathcal{O}}_x|\phi,x\rangle$. The argument is readily generalized to any product of local operators.

\subsection{Correlations in the groundstate}

\begin{figure}
	\includegraphics[scale=1.015]{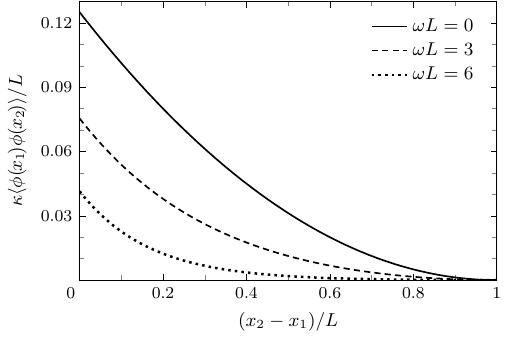}
	\caption{Correlator $\frac{\kappa}{L}\langle\phi(x_1)\phi(x_2)\rangle$ from \eqref{E:<phiphi>}, for a central interval of length $x_2-x_1$, and for different values of $\omega L$. Correlators vanish when $x_2-x_1\to L$ due to the boundary condition $\phi(0)=0=\phi(L)$.}\label{F:centered<phiphi>}
\end{figure}

\begin{figure}
	\includegraphics[scale=1.02]{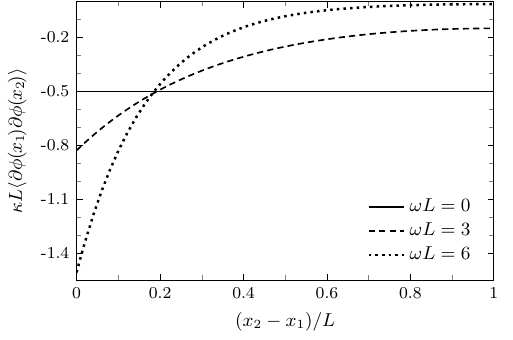}
	\caption{Correlator $\kappa L\langle\partial\phi(x_1)\partial\phi(x_2)\rangle$ from \eqref{E:<SS>},  for a central interval of length $x_2-x_1$, and for different values of $\omega L$.}\label{F:centered<SS>}
\end{figure}

We have shown in Section~\ref{A:separability} that the reduced density matrix $\rho_{A\cup B}$ is a separable mixed state for any disconnected subsystems $A,B$. Therefore, $A$ and $B$ are not entangled, and the correlations left in $\rho_{A\cup B}$ do not arise from entanglement \cite{ollivier2002quantum, giorda2010gaussian,Adesso:2016ygq,adesso2016introduction}. As $A,B$ come into contact, the result does not hold anymore and quantum-driven contact terms are expected. For operators $\hat{\mathcal{O}}_1 , \hat{\mathcal{O}}_2$ local at $x_1 , x_2$, respectively, we define
\beq
\rho(x_1 , x_2)=\int_{-\infty}^{\infty}d\phi_1 d\phi_2 \; f_2 \bigotimes_{i=1,2} |\phi_i , x_i\rangle \langle\phi_i , x_i|\, ,
\eeq
with two-point probability distribution
\beq\label{E:f_2}
\begin{aligned}
&f_2(\phi_1 , \phi_2 ,x_1, x_2)\\
&\quad\;=\mathcal{Z}^{-1}\langle 0,L|\phi_2,x_2\rangle\langle\phi_2,x_2|\phi_1,x_1\rangle\langle\phi_1,x_1|0,0\rangle .
\end{aligned}
\eeq
Then $\langle\hat{\mathcal{O}}_1\hat{\mathcal{O}}_2\rangle = \text{Tr}_{\{x_1, x_2\}}[ \rho (x_1 ,x_2)\hat{\mathcal{O}}_1\hat{\mathcal{O}}_2]$. The function $f_2$ is not symmetric under the \textit{individual} operations $\phi_i\to -\phi_i$. The field correlator and gradient correlator at positions  $x_1<x_2$ are found to be
\begin{align}
\langle\phi(x_1)\phi(x_2)\rangle&=\phantom{-}\frac{\sinh\omega x_1 \sinh\omega (L-x_2)}{2\kappa\omega\sinh\omega L}\, ,\label{E:<phiphi>}\\
\langle\partial_{x_1}\phi(x_1)\partial_{x_2}\phi(x_2)\rangle
	&=-\frac{\omega\cosh\omega x_1 \cosh\omega (L-x_2)}{2\kappa\sinh\omega L}\, .\label{E:<SS>}
\end{align}
The correlators $\frac{\kappa}{L}\langle\phi(x_1)\phi(x_2)\rangle$ and $\kappa L\langle\partial \phi(x_1)\partial \phi(x_2)\rangle$ for a centered interval of length $x_2 - x_1$ are displayed in Figs.\,\ref{F:centered<phiphi>} and \ref{F:centered<SS>} in terms of the dimensionless variable $(x_2-x_1)/L$. Note that these functions depend on $\omega$ and $L$ only through the combination $\omega L$. Because single-point field expectations vanish, we can identify two-point functions $\langle\phi_1 \phi_2\rangle$ and connected correlators $C_{\phi_1,\phi_2}=\langle\phi_1 \phi_2\rangle - \langle\phi_1\rangle\langle\phi_2\rangle$ for this model. By the same token, we identify $\langle\partial\phi_1\partial \phi_2\rangle$ and $C_{\partial\phi_1,\partial\phi_2}=\langle\partial\phi_1 \partial\phi_2\rangle - \langle\partial\phi_1\rangle\langle\partial\phi_2\rangle$.

In the massless limit, $\omega L\to 0$, we recover the results
\begin{align}
\langle\phi(x_1)\phi(x_2)\rangle &= \frac{x_1 (L-x_2)}{2\kappa L}\,,\quad \label{E:phiphi_massless}\\
\langle\partial_{x_1}\phi(x_1)\partial_{x_2}\phi(x_2)\rangle &= -\frac{1}{2\kappa L}\,.
\end{align}
To obtain the thermodynamic limit, the field correlator is best written in the following form:
\beq
\langle\phi(x_1)\phi(x_2)\rangle = \frac{L}{8\kappa}-\frac{x_2 - x_1}{4\kappa}- \frac{(x_1-\frac{L}{2})(x_2-\frac{L}{2})}{2\kappa L}\, .
\eeq
The first term is a divergent contact term or bulk second moment $\langle \phi(L/2)^2\rangle$. The universal, second term is translation-invariant and linearly increasing (in absolute value) with separation, as should be since $\phi$ scales as $(\text{length})^{1/2}$. The final term vanishes when deep enough in the bulk. We thus conclude that the critical Lifshitz boson ($\omega=0$) violates cluster decomposition. Note that $\langle\partial_{x_1}\phi(x_1)\partial_{x_2}\phi(x_2)\rangle_{\text{massless}}$ is constant, and vanishes in the thermodynamic limit, 
\beq
\lim_{L\to\infty}\langle\partial_{x_1}\phi(x_1)\partial_{x_2}\phi(x_2)\rangle_{\text{massless}} = 0 \,,
\eeq
as observed in \cite{chen2017quantum, chen2017gapless} where it meant that bulk $S_z$ spins are uncorrelated in the continuous limit of the groundstate of the Motzkin and Fredkin chains, a result also consistent with the discrete case \cite{movassagh2017entanglement}. 
\red{Taking the spatial derivatives of \eqref{E:phiphi_massless} to get the $\partial_x\phi$ correlator yields a delta-function correlation function (we have assumed non-coincidental points above). The reason for the vanishing of $\langle\partial_{x_1}\phi(x_1)\partial_{x_2}\phi(x_2)\rangle$ deep in the bulk is thus the scaling dimension of $\phi$, which is $-1/2$. As such, one can say that the operator product expansion of $\partial\phi$ with itself only contains an ultra-local contact term.}

Expression \eqref{E:phiphi_massless} confirms that cluster decomposition is not satisfied in the massless case. 
On the other hand, we see from \eqref{E:<phiphi>} that cluster decomposition is satisfied when $\omega >0$ (or $\omega \gg 1/L$ for a finite system). Looking deep into the bulk where $\omega x_1\to\infty$ and $\omega (L-x_2)\to\infty$, we obtain for all values of $\omega(x_2-x_1)$, 
\beq\label{E:phiphi_deep_2}
\langle\phi(x_1)\phi(x_2)\rangle=\frac{1}{4\kappa\omega}e^{-\omega(x_2-x_1)}\,,
\eeq
and
\beq\label{E:dphidphi_deep}
\langle\partial_{x_1}\phi(x_1)\partial_{x_2}\phi(x_2)\rangle=-\frac{\omega}{4\kappa}e^{-\omega(x_2-x_1)}\,,
\eeq
satisfying cluster decomposition at large separations. Connected correlations with an exponential bound $C e^{-m(x_2 - x_1)}$ are characteristic of a gapped phase with mass $m$. We identify $\xi=\omega^{-1}=\sqrt{\kappa}/m$ as the correlation length, and recover the quantum critical Lifshitz boson in the massless limit. Expressions \eqref{E:phiphi_deep_2} and \eqref{E:dphidphi_deep} do not depend on the choice of boundary conditions. We are surprised to find that when the correlation length is infinite (requiring both $\omega\to 0$ and $L\to \infty$ to be effective), the theory develops an IR divergence in bulk correlations. The corresponding limit is singular:
\beq\label{E:corr_IR_1}
\lim_{L\to\infty}\hspace{5pt}\lim_{\omega\to 0}\langle\phi(x_1)\phi(x_2)\rangle \sim \frac{L}{8\kappa}-\frac{x_2-x_1}{4\kappa}\,,
\eeq
and
\beq\label{E:corr_IR_2}
\lim_{\omega\to 0}\hspace{5pt}\lim_{L\to\infty}\langle\phi(x_1)\phi(x_2)\rangle\sim \frac{1}{4\kappa\omega}-\frac{x_2-x_1}{4\kappa}\,,
\eeq
from which we may single out the nonsingular subleading behavior $\langle\phi(x_1)\phi(x_2)\rangle_{\text{regular}}= -(x_2-x_1) / 4\kappa$. A similar phenomenon is observed in the mutual information (see \eqref{E:I_IR_1} and \eqref{E:I_IR_2}).

We emphasize that for $x_1 < x_2$ deep within the bulk,
\beq
\langle\partial_{x_1}\phi(x_1)\partial_{x_2}\phi(x_2)\rangle=\begin{cases}
	\hspace{34pt}0\;, &  m=0\,,\\
	-\frac{\omega}{4\kappa}e^{-\omega(x_2-x_1)}\,, &  m>0\,.
	\end{cases}
\eeq
It is a remarkable fact that, contrary to intuition, this groundstate correlator becomes trivial only for the \textit{massless} theory. Indeed, a mass for $\phi$ corresponds to a gap in the theory, revealed by an exponential decay of correlations in the observables. In the massless limit of large systems, the exponential decay is usually replaced by power-law on separation as the correlation length diverges, resulting in \textit{enhanced} correlations. The vanishing of the self-correlations of an operator as the gap closes is unorthodox.

\subsection{Entanglement in the groundstate}

We consider a finite bipartition $\{A,B\}$ of a one-dimensional system. By analogy with the ``entangling surface'' in higher dimensions, we call ``surface'' the multiple-point boundary between $A$ and $B$, that is $\partial A=\{x_1,x_2,\dots , x_M\}$. Rényi entanglement entropies are defined in \eqref{E:renyi_def}, and we derived in Section \ref{A:replica_trick} an expression
in terms of the $(0+1)$-dimensional partition function $\int_{\text{BC}} \mathcal{D}\phi\; e^{-\int d^d x S_{\text{cl}}[\phi]}$, namely
\beq\label{E:S_n_mass}
S_n (A)=\frac{1}{1-n}\log\Big[\epsilon^{(n-1)M/2}\int d\pmb{\phi}\; \left[f_M(\pmb{\phi}, \mathbf{x})\right]^n\Big]\,,
\eeq
where $f_M (\pmb{\phi} , \mathbf{x})=\mathcal{Z}^{-1}\langle 0 , L | \phi_M , x_M\rangle \cdots \langle \phi_1 , x_1 | 0 , 0\rangle$, and $\epsilon$ is a length scale readily identified as a UV cutoff. In a lattice regularization, for instance, $\epsilon$ is naturally present as the lattice constant. For the massive deformation with Dirichlet conditions, the Gaussian propagators \eqref{E:propagator_mass} give $f_M(\pmb\phi , \mathbf x)=Ce^{-\pmb{\phi}^T K \pmb\phi}$ with normalization factor $C=f_M(\mathbf 0 ,\mathbf x) = \sqrt{\det K / \pi^M}$. Integrating \eqref{E:S_n_mass} over $\mathbb{R}^M$, we find 
\beq\label{E:S_n_mass_gen1}
S_n = -\log \frac{\mathcal{Z}_D^{AB}}{\mathcal{Z}} +\frac{1}2\log\frac{1}{\epsilon^M} - \frac{M}{2(1-n)}\log n \,,
\eeq	
where 
\beq
\frac{\mathcal{Z}_D^{AB} }{ \mathcal{Z}}=\frac{1}{\mathcal{Z}}\langle 0 , L | 0 , x_M\rangle \cdots \langle 0 , x_1 | 0 , 0\rangle=f_M(\mathbf 0,\mathbf x)
\eeq
is a product of propagators with Dirichlet conditions on $\partial A$, normalized by the free propagator $\mathcal{Z}=\langle 0,L|0,0\rangle$. Note that $\mathcal{Z}_D^{AB}=\mathcal{Z}_D^{A}\mathcal{Z}_D^{B}$. We see that $S_n$ is independent of the Rényi index, up to a constant term. We have computed the bipartite Rényi entanglement entropy from first principles, and obtained \eqref{E:S_n_mass_gen1}. This is a generalization to our $(1+1)$-dimensional case of the celebrated Fradkin-Moore formula $S=-\log (Z_D^{AB} / Z_F)$ for the bipartite entanglement entropy of $(2+1)$-dimensional conformal quantum critical theories \cite{fradkin2006entanglement}, which have the RK property and whose groundstates correspond to a lower-dimensional $\text{CFT}_2$. Then $Z_D^{AB}=Z_D^{A}Z_D^{B}$ is the lower-dimensional $\text{CFT}_2$ partition function of configurations with Dirichlet conditions on $\partial A$, and $Z_F$ is the partition function of free configurations. For the massive deformation of the $(1+1)$-dimensional Lifshitz theory, however, the lower-dimensional theory is not conformal invariant, and is only $(0+1)$-dimensional, so that partition functions are simple quantum mechanical propagators. We note that expression \eqref{E:S_n_mass_gen1} at finite mass generalizes naturally to other dimensions.

\begin{figure}
\centering
\includegraphics[scale=1.05]{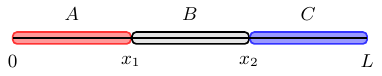}
\caption{Some system partitions. The single-point surface $\{x_1\}$ generates the bipartition $\{A,B\cup C\}$ with Rényi entropies $S_n (A)$ given in \eqref{E:EE_1pt_mass}, while $\{x_1 , x_2 \}$ generates the bipartition $\{A\cup C , B\}$ with Rényi entropies $S_n (B)$ found in \eqref{E:EE_2pt_mass}.}\label{F:ABA}
\end{figure}

\subsubsection{Rényi entanglement entropies}

For a single-point surface separating the ``boundary interval'' $A$ from the rest of the system as in Fig.\,\ref{F:ABA}, we find the Rényi entropies
\beq\label{E:EE_1pt_mass}
S_n (A) =\frac{1}{2}\log\frac{\sinh \omega L_A \sinh\omega (L-L_A)}{\omega\epsilon\sinh \omega L} +b_1(n) \,,
\eeq		
while for a two-point surface separating the bulk interval $B$ from the rest of the system as in Fig.\,\ref{F:ABA}, we obtain
\beq\label{E:EE_2pt_mass}
S_n (B) =\frac{1}{2}\log\frac{\sinh \omega L_A \sinh\omega L_B \sinh\omega L_C}{(\omega\epsilon)^2\sinh \omega L} +b_2(n) \,,
\eeq	 
where the constant $b_M(n)= \frac{M}{2}\log\frac{\pi}{\kappa}- \frac{M}{2(1-n)}\log n$, with $M=1$ and $M=2$ in \eqref{E:EE_1pt_mass} and \eqref{E:EE_2pt_mass}, respectively. The leading universal terms are indeed independent of the Rényi index $n$. These expressions are in perfect agreement with the exact calculation over the discrete versions of the Hamiltonian~\eqref{E:H_Lifshitz_mass} and groundstate~\eqref{E:A_Psi_mass}. Once all interaction terms between a discrete subsystem and its complement have been singled out, one can apply the replica trick and compute a finite number of Gaussian integrals, yielding \eqref{E:S_n_mass_gen1} in the continuous limit, up to constant terms. \red{We apply these results to compute the capacity of entanglement in Appendix \ref{A:capacity}.}

Let us now consider different limiting regimes of the Rényi entropies.
First, in the massless case, expressions \eqref{E:EE_1pt_mass} and \eqref{E:EE_2pt_mass} reduce to
\beq\label{REmassless1}
S_n(A) = \frac{1}2\log\frac{L_A(L-L_A)}{\epsilon L}+ b_1(n)\,,
\eeq
and
\beq\label{REmassless2}
S_n(B) = \frac{1}2\log\frac{L_A L_B L_C}{\epsilon^2 L}+ b_2(n)\,,
\eeq
respectively. The entropy \eqref{REmassless2} of a bulk interval shows an IR divergence, while that of a boundary interval \eqref{REmassless1} does not. Formula \eqref{REmassless1} agrees with that obtained in \cite{angel-ramelli2020logarithmic} for the discrete massless theory, and  matches (B13) of reference \cite{chen2017quantum}. The massless limit of \eqref{E:EE_2pt_mass}, given by \eqref{REmassless2}, however, does not agree with the results of \cite{chen2017quantum}. In fact, our entropy formula \eqref{E:S_n_mass} is different from the one used in \cite{chen2017quantum}.
We will say more about this below \eqref{E:I_massless_Lif}.

\begin{figure}[t]
\includegraphics[scale=1]{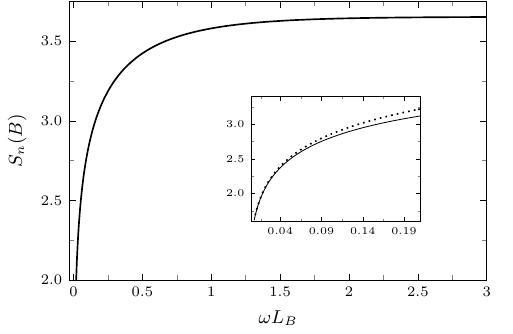}
\caption{Bulk interval Rényi entropy $S_n(B)$ in the large system limit \eqref{E:S_n_bulk}, as a function of $\omega L_B$. The order $n$ and mass-scale $\omega$ are unspecified, contributing only an overall constant. 
An area-law $S_n\sim C_{\text{upper bound}}$ is observed at large values of $\omega L_B$, suppressing dependence on $L_B$  and maintaining only a mass-scale dependence $C_{\text{upper bound}}=\log (\xi/\epsilon)+\text{const.}$, see \eqref{E:EE_long_L_B}. \textbf{Inset:} Small $\omega L_B$ behavior of $S_n$ compared with right-hand side of \eqref{E:S_n_small_mass} (dashed) showing logarithmic deviation from an area-law in the massless case.}\label{F:Sn_mass}
\end{figure}

For large systems we find 
\beq\label{E:S_A_mass_CFT_form}
S_n(A) \underset{L\to\infty}{\longrightarrow} \frac{1}{2}\log\frac{\sinh\omega L_A}{\omega\epsilon} -\frac{1}{2}\omega L_A +\text{const}\,,
\eeq	
and
\beq\label{E:S_n_bulk}
S_n (B) \underset{L_A , L_C\to\infty}{\parbox{1cm}{\rightarrowfill}}\; \frac{1}{2}\log\frac{\sinh\omega L_B}{(\omega\epsilon)^2}  - \frac{1}{2}\omega L_B +\text{const}\,.
\eeq
Here, the constant has the general form $b_M - \frac{Q-1}{2}\log 2$ when $Q$ subintervals have infinite length. At low momenta, quantum fluctuations are suppressed by the effective mass and the groundstate disentangles, meaning that entanglement is localized in the smallest length scales, and only the fine-tuning of mass to zero will entangle the largest scales. A plot of $S_n (B)$ as a function of $\omega L_B$ is given in Fig.~\ref{F:Sn_mass}. In the low mass limit, $\omega L_B\to 0$, the entropy behaves like
\beq\label{E:S_n_small_mass}
S_n (B) = \frac{1}{2}\log \omega L_B  - \frac{1}{2}\omega L_B - \log \epsilon\omega+ \text{const}\,, 
\eeq
departing from the area-law behavior by an unbounded logarithmic dependence on subsystem length, characteristic of a gapless system. 
The prefactor is independent of $n$. In the opposite $\omega L_B\to\infty$ limit (large mass or large subinterval),
\beq\label{E:EE_long_L_B}
S_n (B) =  -  \log \epsilon\omega + \text{const}= \log \frac{\xi}{\epsilon} + \text{const}\,.
\eeq
This is an area law $S_n\sim C_{\text{upper bound}}$ for the gapped system, the general expression for the upper bound being $C_{\text{upper bound}}=\frac{M}{2}\log \frac{\xi}{\epsilon}+\text{const}$, from \eqref{E:S_n_mass_gen1}, where the number of surface points $M$ is the one-dimensional analog of a surface area. The entropy is independent of $L_B$ in this limit, as expected for massive excitations. Recalling that the correlation length $\xi=\omega^{-1} = \sqrt{\kappa}/m$, the upper bound is seen to be a decreasing function of the mass.  

Unexpectedly, we observe a close connection between \eqref{E:S_n_bulk} and the entanglement entropy of a single bulk interval of length $L_B$ (and surface area $M=2$) in a finite temperature $\text{CFT}_2$ \cite{calabrese2004entanglement}, 
\beq
S_1^{\text{CFT}}(L_B,T)=\frac{c}{3}\log \frac{\sinh \pi L_B T}{\pi\epsilon T} + \text{const}\,.
\eeq
With the identification $c=3/2$ and $\pi T=\omega = m/\sqrt{\kappa}$, we can rewrite \eqref{E:S_n_bulk} as
\beq\label{E:S_mass_vs_CFT}
S_n^{\text{Lif}} (L_B,m) = S_1^{\text{CFT}}(L_B,T) - S_1^{\text{CFT}}(L_B,\infty) + \dots\,,
\eeq
where $S_1^{\text{CFT}}(L_B,\infty)=\frac{c}{3}\pi L_B T$ stands for the high temperature asymptotic behavior of $S_1^{\text{CFT}}$, that is, the thermal entropy. The ellipsis stands for nonuniversal terms independent of $L_B$.
The moderately massive Lifshitz theory thus has the entanglement entropy of a moderately hot $\text{CFT}_2$, $S_n^{\text{Lif}} (L_B,m)\sim S_1^{\text{CFT}}(L_B,T)$ for small $T,m$. To our knowledge, only the correspondence between the massless Lifshitz theory and zero temperature $\text{CFT}_2$ had been observed so far.

\subsubsection{Mutual information}

\begin{figure}
\includegraphics[scale=1]{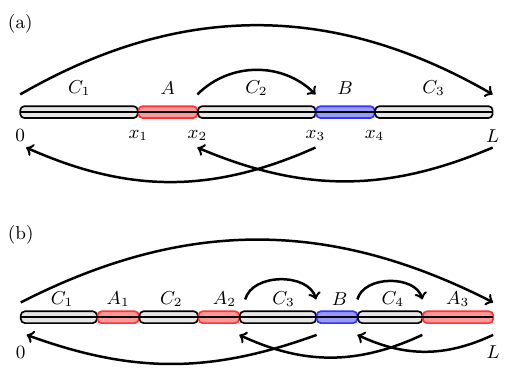}
\caption{Graphical representations of the mutual information between two disconnected bulk intervals $A$ and $B$. In the general expression \eqref{E:I_analytic_gen}, the numerator contains all simple-interval propagators, including one propagator over the full system. Only those not present in the denominator survive, and are represented as arrows above the axis. The remaining denominator propagators are represented as arrows below the axis. The result is a cycle visiting the system's boundary, as well as surface points of one subsystem that are adjacent to a surface point of the other subsystem.}\label{F:CACBC}
\end{figure}

Expression~\eqref{E:S_n_mass_gen1} enables us to give the analytic form of the mutual information $I_n(A:B)=S_n(A)+S_n(B)-S_n(A\cup B)$ between disconnected subsystems $A$ and $B$,
\beq\label{E:I_analytic_gen}
I_n(A:B) = \log\frac{f_{M_A + M_B}(\mathbf 0,\mathbf x_{A}\cup\mathbf x_{B})}{f_{M_A}(\mathbf 0,\mathbf x_{A})f_{M_B}(\mathbf 0,\mathbf x_{B})}\,,
\eeq 
where the $M_A$ surface points of $A$ are positioned at $\mathbf x_{A}$, the $M_B$ surface points of $B$ are positioned at $\mathbf x_{B}$, and 
$\mathbf x_{A}\cap\mathbf x_{B}=\varnothing$. A graphical representation of formula~\eqref{E:I_analytic_gen} for two bulk intervals $A,B$ is given in Fig.\,\hyperref[F:CACBC]{\ref{F:CACBC}(a)}. A graphical example pertaining to another partition is provided in Fig.\,\hyperref[F:CACBC]{\ref{F:CACBC}(b)}, which serves to illustrate a very peculiar property of the (massive) Lifshitz groundstate, namely that the mutual information $I(A:B)$ is insensitive to any component of $A$ with no neighbor in $B$. (In Fig.\,\hyperref[F:CACBC]{\ref{F:CACBC}(b)}, one such component is $A_1$, with no propagators leaving or reaching its surface points.) The mutual information is universal, as expected, and independent of the Rényi index $n$, which will be omitted from now on. (Note that formula \eqref{E:I_analytic_gen} holds whenever a wavefunctional can be mapped to a local action on noncompact fields with propagators that are quadratic forms of the fields, i.e. when $f_M(\pmb\phi , \mathbf x)=f_M(\mathbf 0 ,\mathbf x)e^{-\pmb{\phi}^T K \pmb\phi}$ for multiple-point surface $\partial = \{x_1, \dots , x_M\}$, dimensionless field values $\phi_i \in\mathbb{R}$, and normalization factor $f_M(\mathbf 0 ,\mathbf x) = \sqrt{\det K / \pi^M}$.) For two disconnected bulk intervals as in Fig.\,\hyperref[F:CACBC]{\ref{F:CACBC}(a)} we find
\beq\label{E:I_CACBC}
I(A:B)=\frac{1}{2}\log\frac{\sinh \omega L_{C_1 \cup A \cup C_2}\sinh\omega L_{C_2 \cup B \cup C_3}}{\sinh\omega L_{C_2} \sinh\omega L}\,,
\eeq
which is increasing with the lengths of $A,B$, and decreasing with the separation $L_{C_2}$. In particular, $I(A:B)\to 0$ as $\omega L_{C_2}\to\infty$. In the limit $L_{C_1},L_{C_3}\to\infty$ it becomes
\beq\label{E:I_mass_bulk}
I(A:B)=-\frac{1}{2}\log \sinh\omega L_{C_2} +\frac{1}{2}\omega L_{C_2}-\frac{1}{2}\log 2\,,
\eeq
showing that the dependence on $L_A$ and $L_B$ disappears deep in the bulk, regardless of separation. 
In this limit, we observe a logarithmic divergence at vanishing separations
\beq
I(A:B)\sim\frac{1}{2}\log \frac{1}{\omega L_{C_2}}\,, \qquad \omega L_{C_2}\to 0\,,
\eeq
and an exponential decay at large separations
\beq
I(A:B)\sim\frac{1}{2}e^{-2\omega L_{C_2}}\,, \qquad \omega L_{C_2}\to \infty\,.
\eeq
However, the mutual information does not vanish when the size of $A$ and/or $B$ vanishes. This is in stark contrast with most common theories. In a free fermion $\text{CFT}_2$, for instance, the mutual information between two intervals of length $\ell$ separated by a distance $r$ is $I(\ell,r)=\frac{1}{3}\log\frac{(r+\ell)^2}{r(2\ell +r)}$ \cite{Casini:2005rm,swingle2010mutual}, which clearly vanishes as $\ell$ tends to zero.
We note that in the massless limit of \eqref{E:I_CACBC}, we obtain
\beq\label{E:I_massless_Lif}
I(A:B)=\frac{1}{2}\log \frac{(L_{C_1}+L_A +L_{C_2})(L_{C_2} + L_B + L_{C_3})}{L_{C_2}L}\,,
\eeq
which does not agree with the result found in \cite{chen2017quantum} where mutual information seems to vanish. In this reference, the reduced density $\rho_{A\cup B}$ is assumed, wrongly, to depend only on the difference fields $\phi_A \equiv \phi_2 - \phi_1$ and $\phi_B \equiv \phi_4 - \phi_3$, where $A,B$ are as in Fig.\,\hyperref[F:CACBC]{\ref{F:CACBC}(a)} and $\phi_i = \phi(x_i)$. The resulting expression is of the form $\rho_{A\cup B}\sim \int \mathcal{D}\phi_A \mathcal{D}\phi_B \, h(\phi_A , \phi_B)|\phi_A \phi_B\rangle\langle \phi_A \phi_B |$, with $h$ a Gaussian function of $\phi_A ,\phi_B$. This expression, however, neglects the crucial dependence of $\rho_{A\cup B}$ on the ``anchor" $\pmb\phi = (\phi_1 , \phi_2 , \phi_3 , \phi_4)$, as derived in~\eqref{E:separable_appendix}. A similar comment holds for other reduced densities as well. We have already mentioned that the expression obtained in \cite{chen2017quantum} for the entanglement entropy of a single bulk interval is at odds with (the massless limit of) our expression~\eqref{E:EE_2pt_mass}. We find that far from vanishing, the mutual information diverges like $\frac{1}{2}\log L/L_{C_2}$ deep in the bulk. This is not entirely unexpected since correlators of the massless theory do not possess the cluster decomposition property. It is another manifestation of the bulk IR divergence that develops when the correlation length goes infinite (requiring both $\omega\to 0$ and $L\to \infty$ to be effective). The corresponding limit is singular:
\beq\label{E:I_IR_1}
\lim_{L_{C_1},L_{C_3}\to\infty}\hspace{5pt}\lim_{\omega\to 0} I(A:B) \sim \frac{1}{2}\log L-\frac{1}{2}\log L_{C_2}\,,
\eeq
and
\beq\label{E:I_IR_2}
\lim_{\omega\to 0}\hspace{5pt}\lim_{L_{C_1},L_{C_3}\to\infty} I(A:B) \sim \frac{1}{2}\log\frac{1}{\omega}-\frac{1}{2}\log L_{C_2}\,,
\eeq
which to be compare with the divergence in field correlations in \eqref{E:corr_IR_1} and \eqref{E:corr_IR_2}.

We note that the mutual information shared by two boundary intervals, i.e.~taking $L_{C_1}=L_{C_3}=0$ in \eqref{E:I_massless_Lif}, reduces to%
\footnote{ Coincidentally, the mutual information \eqref{E:MI_bd} shared by two boundary intervals for the massless Lifshitz theory takes the exact same form (with a prefactor of $3/2$) as the mutual information between two intervals separated by a distance $r$ in an infinite system for free fermion CFT$_2$ \cite{Casini:2005rm,swingle2010mutual},
$$I_{\text{free fermions}}(A:B)=\frac{1}{3}\log\frac{(L_A+r)(L_B+r)}{r(L_A+L_B +r)}\,,$$ noticing that $L=L_A+L_B+r$, with $r\equiv L_{C_2}$ in \eqref{E:MI_bd}.}
\beq\label{E:MI_bd}
I(A:B) = \frac{1}2 \log \frac{(L-L_A)(L-L_B)}{L(L-L_A-L_B)}\,,
\eeq
since $L=L_A+L_B+L_{C_2}$. Then, in the large $L$ limit (or equivalently large separation), the mutual information does vanish as
\beq
I(A:B)\xrightarrow[L\to \infty]{}\frac{L_AL_B}{2L^2}\,.
\eeq
The vanishing of the mutual information between two far apart boundary intervals can be understood by looking at the fields correlator  \eqref{E:phiphi_massless}. For this configuration, the two surface points separating $A\cup B$ from the rest of the system are located at $x_1=L_A$ and $x_2=L-L_B$, and the corresponding correlator vanishes in the regime $L\gg L_A,L_B$, i.e.\,\,$\langle\phi(x_1)\phi(x_2)\rangle = L_AL_B/(2\kappa L)\rightarrow 0$, hence recovering cluster decomposition.

Finally, we may also compute the mutual information between two adjacent subsystems $A,B$ (see Fig.\,\hyperref[F:CACBC]{\ref{F:CACBC}(a)} setting $L_{C_2}$ empty) using the entropy \eqref{E:EE_2pt_mass} for an interval in the bulk. One finds
\beq
\begin{aligned}\label{MI_adj}
&I(A:B)\\
&\;\;\,=\frac{1}{2}\log\frac{\sinh \omega L_A \sinh\omega L_B \sinh\omega L_{C_1\cup A}\sinh\omega L_{B\cup C_3}}{(\omega\epsilon)^2\sinh \omega L\sinh \omega L_{A\cup B}} ,
\end{aligned}
\eeq
where we omitted an unimportant constant. It presents, as expected, a UV divergence. Deep in the bulk, i.e.~in the limit $L_{C_1},L_{C_3}\to\infty$, the mutual information shared by two adjacent subsystems behaves as
\beq
\begin{aligned}\label{MI_adj_inf}
&I(A:B)\sim\frac{1}{2}\log\frac{\sinh \omega L_A \sinh\omega L_B}{(\omega\epsilon)^2\sinh \omega L_{A\cup B}} \,,
\end{aligned}
\eeq
dependent only on $L_A$ and $L_B$. The massless limit of \eqref{MI_adj} reads
\beq
\begin{aligned}\label{MI_adj_massless}
I(A:B)=\frac{1}{2}\log\frac{(L_{C_1}+L_A)L_A  L_B (L_B+L_{C_3})}{\epsilon^2( L_A +L_B)L} \,,
\end{aligned}
\eeq
which, deep in the bulk, yields
\beq
\begin{aligned}\label{MI_adj_massless2}
I(A:B)\sim\frac{1}{2}\log\frac{L_A  L_B}{\epsilon( L_A +L_B)} +\frac{1}{2}\log\frac{L}\epsilon\,.
\end{aligned}
\eeq
Taking the massless limit $\omega L_{A/B}\rightarrow0$ from \eqref{MI_adj_inf} instead, one gets
\beq
\begin{aligned}\label{MI_adj_massless3}
I(A:B)\sim\frac{1}{2}\log\frac{L_A  L_B}{\epsilon(L_A +L_B)} -\frac{1}{2}\log\omega\epsilon\,.
\end{aligned}
\eeq
We thus observe both UV and IR divergences in the mutual information between adjacent subsystems. The first term is of the same form as the mutual information and logarithmic negativity between two adjacent intervals in an infinite system for a CFT$_2$~\cite{Calabrese:2012ew}.


\subsubsection{Entropic $c$-function}

\begin{figure}[t]
	\includegraphics[scale=1]{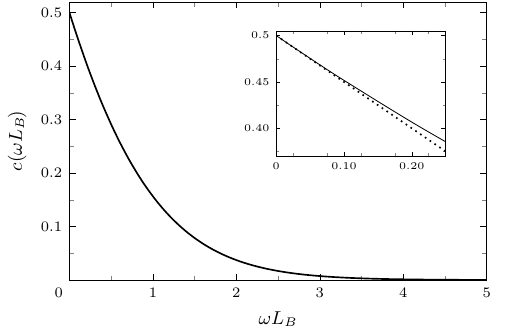}
	\caption{Bulk interval $c$-function, see \eqref{E:cfct_mass}, as a function of dimensionless subsystem mass-length $\omega L_B$. \textbf{Inset:} Small $\omega L_B$ behavior of $c$ compared with right-hand side of \eqref{E:cfct_small_mass}, first line (dashed). The curve tends to the critical value $c=1/2$.}\label{F:cn_mass}
\end{figure}

Being universal, i.e.~cutoff independent, $c$-functions are important tools in the study of (unitary) Lorentz-invariant field theories and their RG fixed points, and are known to be monotone decreasing for such theories in two dimensions \cite{Zamolodchikov1986irreversibility, casini2004finite}. The decreasing of the entanglement entropy under the RG flow is a property demonstrably true for relativistic theories \cite{casini2004finite}, but known to admit nonrelativistic exceptions \cite{laflorencie2015quantum}. The entropic $c_n$-function is defined as
\beq
c_n(\ell)= \ell \frac{dS_n}{d \ell}\,,
\eeq
where $S_n$ is the (Rényi) entanglement entropy of an interval of length $\ell$. Regrettably, $c$-functions are difficult to compute for general theories, and analytical answers are scarce and most explicit results are obtained numerically. Here, however, we are able to analytically compute the entropic $c$-function of the massive Lifshitz theory.

The entropic $c_n$-function corresponding to \eqref{E:S_n_bulk} is then
\beq\label{E:cfct_mass}
c (\omega L_B)=\frac{1}{2}\omega L_B(\coth \omega L_B - 1)\,,
\eeq
with asymptotics
\beq\label{E:cfct_small_mass}
c (\omega L_B) \sim \begin{cases}
\displaystyle\frac{1}{2} -\frac{1}{2}\omega L_B\,, & \quad\omega L_B \ll 1 \,,\vspace{5pt}\\
\omega L_B\, e^{-2\omega L_B}\,, & \quad\omega L_B\gg 1 \,.
\end{cases}
\eeq
\noindent Note that we have dropped the $n$-dependence since it plays no role in this theory. A plot is given in Fig.\,\ref{F:cn_mass}. Interestingly, it is monotone decreasing under wavefunction RG flow, even though the theory is \textit{not} Lorentz invariant. (In fact, not even Lifshitz invariant.) Lorentz invariance is thus not necessary for the monotone decreasing of the $c$-function. Identifying the necessary conditions is still an open question.


\subsection{2D corner entanglement by dimensional reduction}\label{S:2+1_cornerEE}

\begin{figure}\vspace{-15pt}
\includegraphics[scale=1.05]{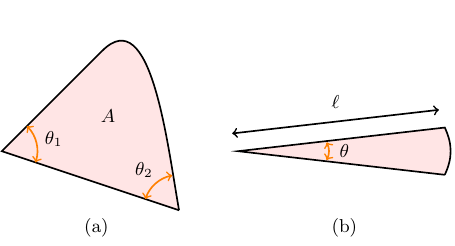}
\caption{(a) Subsystem $A$ with two corners in the one-dimensional boundary, with opening angles $\theta_1$ and $\theta_2$. (b)~Sharp corner characterized by the linear dimension $\ell$.}\label{F:corner_entanglement}
\end{figure}

With a few important exceptions, quantum critical states in $2+1$ spacetime dimensions have Rényi entropy that scales as
\beq\label{E:Sn_corner_generic}
S (A) = \mathcal{B} \frac{\ell}{\epsilon} - \sum_i a (\theta_i)\log\frac{\ell}{\epsilon} + \text{const}\,,
\eeq
where $\ell$ is a linear dimension characterizing subregion $A$, $\epsilon\ll\ell$ is a UV cutoff, and $i$ ranges over corners in the one-dimensional boundary of $A$, each with opening angle $\theta_i$ (see Fig.\,\ref{F:corner_entanglement} (a)). The subleading logarithmically divergent terms measure the so-called \textit{corner entanglement}, and the corner function $a(\theta)$ is universal and expected to depend only on scale-invariant geometric features of $A$ \cite{casini2006universal, casini2008entanglement, fradkin2013field}.

In the sharp corner limit $\theta\to 0$ of non-interacting CFTs, the corner function may be analytically computed through dimensional reduction, by first considering a rectangular strip $A$ with length $\ell$ and width $r\ll\ell$. The corresponding entropy is \cite{Casini2009entanglement}
\beq\label{E:Sn_strip}
S (A)=\mathcal{B} \frac{\ell}{\epsilon} -\kappa_c\frac{\ell}{r}\,,
\eeq
where $\kappa_c$ is a universal coefficient also characterizing the mutual information shared by two regions $A,B$ of length $\ell$ and separation $r\ll\ell$:
\beq\label{E:I_strip}
I(A:B) = \kappa_c \frac{\ell}{r}\,.
\eeq
Expressions \eqref{E:Sn_strip} and \eqref{E:I_strip} are examples of relations that hold when a region of interest is characterized by $m$ large linear dimensions $\ell_1, \dots , \ell_m$ and $n$ small linear dimensions $r_1 , \dots , r_n$. Indeed, the entropy of such a region should be extensive in the large dimensions $\ell_i$ which may then be considered periodic and Fourier analyzed, effectively reducing the dimension from $m+n$ to $n$. In the case of the thin rectangular strip $A$, the dimensional reduction leaves only one space dimension, and recasts $\kappa_c$ into an integral of the entropic $c$-function of the lower dimensional QFT \cite{Casini2009entanglement}: 
\beq\label{E:kappa_integral_cn}
\kappa_c = \frac{1}{\pi}\int_0^{\infty} dx\, c (x) \,.
\eeq
A sharp corner of length $\ell$ and opening angle $\theta\ll 1$, as in Fig.\,\ref{F:corner_entanglement} (b), can then be built out of thin strips of length $d\ell'$ and width $\theta \ell'$, such that  from~\eqref{E:Sn_strip} one obtains
\beq
S(A)=\int_{\epsilon}^{\ell} d\ell' \left(\frac{\mathcal{B}}{\epsilon}-\frac{\kappa_c}{\theta\ell'}\right)=\mathcal{B} \frac{\ell}{\epsilon} -\frac{\kappa_c}{\theta}\log\frac{\ell}{\epsilon}\,.
\eeq
Comparison with~\eqref{E:Sn_corner_generic} gives
\beq
a(\theta\to 0)=\frac{\kappa_c}{\theta}\,.
\eeq
As it characterizes the (regulator-independent) leading divergence of $a(\theta\to 0)$, $\kappa_c$ is also called the \textit{sharp limit coefficient}.

We now consider a \textit{nonconformal} theory, namely the $(2+1)$-dimensional $z=2$ real Lifshitz theory, with Lagrangian density
\beq
\mathcal{L}_{2} = \frac{1}{2}\left[\dot\phi^2 - \kappa^2 \left(\nabla^2\phi\right)^{2}\right].
\eeq
Compactifying $y$ to an $S_1$ of circumference $\ell_y$, we decompose the field and the Lagrangian in Fourier modes of momenta $k=2\pi n/\ell_y$, $n\in\mathbb{Z}$:
\beq
\begin{aligned}
\phi(x,y)&=\sum_{k} e^{iky}\phi_k (x)\,,\\
\mathcal{L}_{2}(x,y)&=\sum_{k,k'} e^{i(k'+k)y}\mathcal{L}_{2,k,k'}(x)\,,
\end{aligned}
\eeq
with
\beq
\mathcal{L}_{2,k,k'}= \frac{1}{2}\left[\dot\phi_k\dot\phi_{k'} - \kappa^2 \phi_k\left(\partial_x^2 - k^2\right)\left(\partial_x^2 -k'^2\right)\phi_{k'}\right].
\eeq
Integrating $y$ out, we obtain $\mathcal{L}_{1}(x)=\int_{S_1} dy \,\mathcal{L}_{2}=\ell_y \sum_k \mathcal{L}_{1,k}$ where
\beq
\mathcal{L}_{1,k}=\frac{1}{2}\left[|\dot\phi_k|^2 - \kappa^2 |\partial_x^2 \phi_k |^2 -2\kappa m_k^2 |\partial_x \phi_k |^2 - m_k^4 |\phi_k |^2\right],
\eeq
and $m_k^2 = \kappa k^2$. Since $\mathcal{L}_{1,k}(\phi_k)=\mathcal{L}_{1,k}(\text{Re }\phi_k)+\mathcal{L}_{1,k}(\text{Im }\phi_k)$, each decoupled mode $\phi_k$ is a double copy of the real $(1+1)$-dimensional massive deformation with Hamiltonian~\eqref{E:H_Lifshitz_mass}. Notice the fine-tuning of the $|\partial_x \phi_k |^2$ term, just as in~\eqref{E:H_Lifshitz_mass}, a direct consequence of the quadraticity of the Hamiltonian $\int d^d x\, A^{\dagger}A$. We find from the corresponding lower dimensional $c$-function~\eqref{E:cfct_mass}, and identity~\eqref{E:kappa_integral_cn},
\beq\label{integC}
\kappa_c = \frac{1}{\pi}\int_0^{\infty} dx\, \frac{x}{2}(\coth x -1) = \frac{\pi}{24}\,.
\eeq
This result is in agreement with previous results pertaining to the $(2+1)$-dimensional $z=2$ Lifshitz theory where, owing to the \textit{spatial} conformal invariance of the groundstate wavefunctional, the corner function may be given in closed-form \cite{fradkin2006entanglement}. It is remarkable that relationship~\eqref{E:kappa_integral_cn} between $\kappa_c$ and the entropic $c$-function still holds for the $z=2$ Lifshitz theory, although its scale invariance is Lifshitz (anisotropic) instead of conformal. Interestingly, there is compelling evidence that the corner function can serve as a measure of the number of low-lying degrees of freedom in theories near criticality \cite{fradkin2006entanglement, kallin2014corner, bueno2016bounds}, generalizing the role played by the central charge of $(1+1)$-dimensional CFTs.


\section{Positive boson and Motzkin and Fredkin chains}\label{S:positive}

In this section, we consider the positive-field version of the (massless) Lifshitz groundstate \eqref{E:Psi_0}.  Deep inside the bulk, the constraint $\phi\ge0$ can be ignored such that correlation and entanglement properties are identical to those of the unconstrained theory. We compute the Rényi entropies of a boundary interval and of a bulk interval as an illustration. Finally, we comment on the relation between the positive Lifshitz boson and the Motzkin and Fredkin groundstates, and discuss results in the literature.

\subsection{Positive Lifshitz boson}
The positive-field version of the groundstate \eqref{E:Psi_0} reads
\beq\label{E:Psi_pos}
\Psi_0[\phi]=\frac{1}{\sqrt{\mathcal{Z}}}\hspace{1pt}e^{-\frac{1}{2}\int d^d x\, \kappa (\nabla\phi)^2} \prod_x \theta(\phi(x))\,,
\eeq
where $\theta(z)$ is the Heaviside function that enforces $\phi$ to be non-negative. This wavefunctional was conjectured in \cite{chen2017quantum,chen2017gapless} to be a continuum version of the groundstate of both Motzkin and Fredkin spin chains. The $\phi\ge0$ constraint is then applied to match that of the lattice
Motzkin/Dyck paths that appear in the groundstate.

The corresponding propagator can be expressed in terms of the free one \eqref{E:propagator_Lif} as
\beq
\begin{aligned}
G_{\mathbb{R}^{+}}(\phi' , x' ; \phi , x)=G_{\mathbb{R}}(\phi' , x' ; \phi , x)-G_{\mathbb{R}}(\phi' , x' ; -\phi , x)\,,
\end{aligned}
\eeq
which explicitly reads
\beq
\begin{aligned}
&G_{\mathbb{R}^+}(\phi' , x' ; \phi , x)\\
&\qquad=\sqrt{\frac{\kappa}{\pi (x'-x)}}\bigg(e^{- \tfrac{\kappa(\phi'-\phi)^2}{x'-x}}-e^{-\tfrac{\kappa(\phi'+\phi)^2}{x'-x}}\bigg)\,.\label{prop_massless_pos}
\end{aligned}
\eeq
By construction, the propagator $G_{\mathbb{R}^+}(\phi' , x' ; \phi , x)$ vanishes for $\phi=0$ or $\phi'=0$. Upon setting Dirichlet boundary conditions at $x=0,L$, one thus needs to introduce a regulator $h$, i.e.~$\phi(0,t)=\phi(L,t)=h$, which will be sent to zero at the end of the calculations.
Furthermore, as the $M(>\hspace{-2pt}1)$-point functions for the positive boson are not Gaussian (they are sums of Gaussians), one cannot apply the (generalized) Fradkin-Moore formula \eqref{E:S_n_mass_gen1}. Instead, we compute the Rényi entropies directly from \eqref{E:S_n_mass}.

For a boundary interval $A$, as in Fig.\,\ref{F:ABA}, it is straightforward to find
\beq\label{E:EE_pos1}
S_n (A) =\frac{1}{2}\log\frac{L_A(L-L_A)}{\epsilon L} +b(n) \,,
\eeq		
where $b(n)=\hspace{-1pt}\tfrac{2n+1}{2(n-1)}\hspace{-1pt}\log\hspace{-1pt} n + \tfrac{1}{1-n}\hspace{-1pt}\log\hspace{-1pt} \tfrac{\Gamma\left(n+1/2\right)}{\Gamma (3/2)^n}-\tfrac{1}{2}\hspace{-1pt}\log4\kappa$.

For an interval $B$ in the bulk (see Fig.\,\ref{F:ABA}), the moments of the associated reduced density matrix are given by the following complicated expression in terms of the hypergeometric function:
\beq
\begin{aligned}
&{\rm Tr}\,\rho_B^n\\
&\;\;= \frac{1}\pi\Big(\frac{n\pi}{4\kappa }\Big)^{1-n}\bigg(\hspace{-1pt}\frac{L_AL_BL_C}{\epsilon L}\hspace{-1pt}\bigg)^{(1-n)/2}w^{n/2}(w-1)^{n+1/2}\\
&\quad\;\;\times\sum_{k=0}^n(-1)^k\binom{n}{k}\bigg[\tfrac{1}{n^2}\Gamma\big(\hspace{-1pt}\tfrac{n+1}2\hspace{-1pt}\big)^2\, _2F_1\big(\hspace{-1pt}\tfrac{n+1}2,\tfrac{n+1}2,\tfrac{1}2,\tfrac{a_k^2}w \hspace{-1pt}\big)\\
&\qquad\; + \tfrac{\sqrt{w}\Gamma(n/2)^2}{2(n+1)^2a_k}\Big(\hspace{-1pt}(w^{-1}a_k^2-1)\, _2F_1\big(\hspace{-1pt}\tfrac{n+2}2,\tfrac{n+2}2,\tfrac{-1}2,\tfrac{a_k^2}w\hspace{-1pt}\big)\\
&\qquad\; + (1-2(n+2)a_k^2w^{-1})_2F_1\big(\hspace{-1pt}\tfrac{n+2}2,\tfrac{n+2}2,\tfrac{1}2,\tfrac{a_k^2}w\hspace{-1pt}\big)\Big)\bigg]\,,
\end{aligned}
\eeq
where $w=(L_A + L_B)(L_B + L_C)/(L_A L_C)$, and $a_k=1-2k/n$.
However, deep in the bulk, i.e. for $w\rightarrow1$, only the $k=0,n$ terms contribute to leading order in the sum above. We find
\beq\label{E:EE_pos2}
\begin{aligned}
&{\rm Tr}\,\rho_B^n\\
&\quad\simeq \frac{1}\pi\Big(\frac{n\pi}{4\kappa }\Big)^{\hspace{-1pt}1-n}\bigg(\hspace{-1pt}\frac{L_AL_BL_C}{\epsilon L}\hspace{-1pt}\bigg)^{\hspace{-2pt}(1-n)/2}w^{n/2}(w-1)^{n+1/2}\\
&\qquad\qquad\times\bigg[n^{-2}2\sqrt{\pi}\Gamma\big(\hspace{-1pt}n+\tfrac{1}2\hspace{-1pt}\big)\frac{1}{(w-1)^{n+1/2}}+\cdots\bigg]\\
&\quad=\bigg(\hspace{-1pt}\frac{L_AL_BL_C}{\epsilon L}\hspace{-1pt}\bigg)^{\hspace{-1pt}(1-n)/2}\Big(\frac{n\pi}{4\kappa }\Big)^{\hspace{-1pt}1-n}\hspace{-1pt}\frac{2}{n^2\sqrt{\pi}}\Gamma\big(\hspace{-1pt}n+\tfrac{1}2\hspace{-1pt}\big)+\cdots,
\end{aligned}
\eeq
and one can check that this expression is properly normalized for $n=1$.
Finally, the Rényi entropy reads
\beq
\label{E:EE_pos2b}
\begin{aligned}
S_n (B)&=\frac{1}{2}\log\frac{L_A L_B L_C}{\epsilon^2 L}+ {\rm const} \,,\\
&\sim\frac{1}{2}\log\frac{L_B}{\epsilon} +\frac{1}{2}\log\frac{L}{\epsilon}
\end{aligned}
\eeq
valid for $L_A,L_C\sim L\gg L_B$. We thus retrieve the entropy of the free massless theory in the regime where $B$ is deep enough in the bulk. This is actually a general feature; the constraint $\phi\ge 0$ is inconsequential deep inside the bulk due to the exponentially small probability of $\phi$ approaching zero.
Thus, in that regime, we expect correlation functions\footnote{The (connected) correlation functions for the positive Lifshitz boson may be found in Appendix \ref{S:singular}, setting $\nu=1/2$ in the corresponding expressions.} and Rényi entropies to be given by those for the unconstrained boson (see Section \ref{S:mass}). In particular, the mutual information of two disconnected intervals deep in the bulk is given by \eqref{E:I_IR_1}, i.e.
\beq\label{MI_dis}
I(A:B)\sim \frac{1}2\log\frac{L}{L_{C_2}}\,,
\eeq
and that of two adjacent intervals by \eqref{MI_adj_massless2}, that is
\beq\label{MI_adj2}
I(A:B)\sim\frac{1}{2}\log\frac{L_AL_B}{\epsilon(L_A+L_B)}+\frac{1}{2}\log\frac{L}{\epsilon}\,.
\eeq
The former is only IR divergent while the latter is both UV and IR divergent.

\subsection{Relation to Motzkin and Fredkin chains}

\red{The Motzkin and Fredkin models are spin-1 and spin-$1/2$ chains, respectively. The Hamiltonians describing these spin chains are local, frustration-free, and with nearest-neighbor interactions.
Their groundstates are equal-weight superpositions of all states corresponding to Motzkin paths for the Motzkin model, and to Dyck paths for the Fredkin model. A Motzkin path is any path connecting $(x, y) = (0, 0)$ and $(x, y) = (L, 0)$ in the upper-half plane, formed by three types of moves (moving from left to right): diagonal-up, diagonal-down, and flat, (which correspond to the three state in the $S_z$ basis for spin-1). Similarly, a Dyck path is a Motzkin path with only the diagonal-up/down moves allowed.
The spin variables can thus be represented as a ``height field" $\phi$, i.e. $\partial_x \phi(x)=S_z(x)$. The groundstate property $S_z^{\text{tot}}=0$ translates into Dirichlet boundary conditions on $\phi$: $\phi(L)-\phi(0)=0$. Mapping the problem to a random walk, 
}%
the wavefunctional \eqref{E:Psi_pos} of the positive Lifshitz boson was thus conjectured in \cite{chen2017quantum,chen2017gapless} to capture the groundstates of both Motzkin and Fredkin spin chains in the continuum limit. The only difference between the two being encoded in the dimensionless parameter $\kappa$. Indeed, expression \eqref{E:EE_pos1} for the Rényi entropy of a boundary interval matches exactly the results for the Motzkin model \cite{movassagh2017entanglement,chen2017quantum} and for the Fredkin spin chain \cite{DellAnna2016violation,chen2017gapless}, provided $\kappa = 3/4,\,2$ in the Motzkin and Fredkin groundstates, respectively. Correlation functions can be shown to match as well. However, our field theory result \eqref{E:EE_pos2b} for the (Rényi) entropy of a bulk interval does not agree with the spin chains calculations \cite{movassagh2017entanglement,dellanna2019long}. Our formula does reproduce the `geometric' part (i.e. the $1/2\log L_B$ term) of the entropy of Motzkin and Fredkin groundstates, but we also observe an additional IR divergent term that does not appear for the spin chains. We are thus lead to conclude that, as it stands, the continuum limit of Motzkin and Fredkin groundstates is not adequately described by the positive Lifshitz boson groundstate. In the Discussion, we discuss one modification to the wavefunctional that could potentially cure the discrepancies, namely making the field compact.

Let us close this section with a remark on the mutual information for the Motzkin and Fredkin groundstates, which was studied in \cite{dellanna2019long}. There, the author finds that the mutual information of two disconnected subsystems inside the bulk does not depend on their separation, and takes the form
\beq
\begin{aligned}\label{MI_dellanna}
I(A:B)\sim\frac{1}{2}\log\frac{L_A  L_B}{L_A +L_B}+ {\rm const}\,,
\end{aligned}
\eeq
in the limit $L_A,L_B\gg1$, where $L_X$ is the number of spins in the subsystem $X$. Surprisingly, this formula coincides with the geometric term (first term) in \eqref{MI_adj2}, which gives the mutual information shared by two \textit{adjacent} intervals for the Lifshitz groundstate. Furthermore, a length scale seems to be missing in \eqref{MI_dellanna} if one wishes to take the continuum limit. Indeed, expression \eqref{MI_dellanna} would then appear to be UV divergent, which is in contradiction with the fact that mutual information between disconnected subsystems must be UV finite.

\section{Discussion and outlook}\label{S:conclu}

In this work, we considered an important and vast class of quantum states, namely Rokhsar-Kivelson states that quantum mechanically encode the partition function of a classical system. We studied entanglement and correlations properties of {\slshape continuum} RK states for which the dual classical models are local QFTs. As instances of such states, Lifshitz groundstates and their deformations were analyzed.

We proved in Section~\ref{A:separability} that for real-valued RK wavefunctionals of ($d+1$)-dimensional QFTs, the reduced density matrix $\rho_{A \cup B}$ of two disconnected subsystems $A$ and $B$ is (mixed) separable, meaning that the partial trace over $C$ (the complement of $A \cup B$) disentangles $A$ from $B$. Accordingly, the mutual information results entirely from classical and quantum non-entangling correlations \cite{ollivier2002quantum, giorda2010gaussian,Adesso:2016ygq,adesso2016introduction}.
Furthermore, the separability of $\rho_{A \cup B}$ for two disjoint subsystems implies the vanishing of the logarithmic negativity. This is in agreement with previous results where the logarithmic negativity was shown to vanish; in \cite{chen2017gapless} for the groundstate of the noncompact free massless $z=2$ Lifshitz boson, and in \cite{angel-ramelli2020logarithmic} for local groundstates of free massless Lifshitz theories with even positive integer $z$, both for compact and noncompact fields.

We have introduced nontrivial deformations of the noncompact $z=2$ Lifshitz theory that preserve the RK structure of the original theory. Accordingly, the correlators of the groundstate wavefunctional are encoded in the partition function of a lower-dimensional, local action. For the massive deformation studied in Section~\ref{S:mass}, which explicitly breaks Lifshitz scaling, groundstate correlations are given by the Euclidean harmonic oscillator. As expected, we found an exponential decay of correlations with correlation length $\xi\sim m^{-1}$, and observed that cluster decomposition, violated in the massless case, is restored by the regulating mass. 

We further computed, from first principles, the Rényi entanglement entropy over a general bipartition $\{A,B\}$ with finitely many surface points, $\partial A = \{x_1, \dots , x_M\}$, and the corresponding $c$-function (independent of the Rényi index $n$). Expression \eqref{E:S_n_mass_gen1} for $S_n (A)$ generalizes to our $(1+1)$-dimensional non-Lifshitz-invariant case the celebrated Fradkin-Moore formula \cite{fradkin2006entanglement} pertaining to the $(2+1)$-dimensional $z=2$ Lifshitz theory. Application of this formula yields a general expression for the UV-finite mutual information between disjoint subsystems $A_1$ and $A_2$. When $A_1 , A_2$ are two intervals, the mutual information decays exponentially with separation, unless the mass is zero. In the massless case, the mutual information is IR divergent and persists at all separations, which can be seen as a consequence of the cluster decomposition being violated.

We also observed a relationship between corner entanglement in the $(2+1)$-dimensional $z=2$ Lifshitz theory and the entropic $c$-function of the $(1+1)$-dimensional massive deformation. Specifically, the sharp-limit coefficient $\kappa_c$ of the corner entanglement is found to be the integrated $c$-function, see \eqref{integC},
for our non-Lorentz-invariant theory. This is a generalization of the corresponding relation between the sharp-limit coefficient of a $\text{CFT}_3$ and the entropic $c$-function of a lower-dimensional QFT \cite{Casini2009entanglement}. Interestingly, the entropic $c$-function is found to decrease under wavefunction RG flow even though the theory is not Lorentz-invariant. (In fact, not even Lifshitz-invariant.) Whether a similar result may hold for higher dimensional versions of the massive deformation presented in this work is worth investigating, especially in dimension $2+1$, where the $F$-theorem for CFTs guaranties the monotone decreasing of the entanglement entropy of a disk under RG flow \cite{Casini2012renormalization}. We leave this for future study.

As a byproduct of our results on Rényi entropies, we computed the capacity of entanglement $C_A$ in Appendix~\ref{A:capacity}. This quantity, analog of the heat capacity for thermal states, characterizes the width of the entanglement spectrum. We find that for Lifshitz groundstates, the capacity of entanglement is finite and follows an area law. Comparison of capacity of entanglement with entanglement entropy tells us something about the entanglement structure of the quantum state under study. We found $C_A\ll S_A$ for Lifshitz groundstates, which could be interpreted as entanglement being effectively carried by maximally entangled EPR pairs in such systems.

One motivation for starting this work was in relation to the Motzkin and Fredkin spin chains \cite{Bravyi2012criticality,DellAnna2016violation}. The positive-field version of the Lifshitz groundstate \eqref{E:Psi_0} was believed to be the continuum limit of the Motzkin and Fredkin groundstates, and, indeed, captures many of their spin and entanglement features \cite{chen2017quantum, chen2017gapless, movassagh2017entanglement}. 
In comparing our field theory predictions to the spin chains results in the literature, the entanglement entropy of a bulk interval \eqref{REmassless2} obtained here does not agree with that computed in \cite{movassagh2017entanglement}.
An extra IR piece, absent for the spin chains, is at the origin of the discrepancy. This also implies that the form of the mutual information \eqref{MI_dis} cannot be that for Motzkin and Fredkin groundstates.
Indeed, the mutual information \eqref{MI_dis} for disconnected intervals obtained here does not agree with that of \cite{dellanna2019long} computed on the lattice. However, as argued in Section \ref{S:positive}, the expression found in \cite{dellanna2019long} (see \eqref{MI_dellanna}) does not possess a sensible continuum limit. Still, as it stands, the positive Lifshitz boson cannot be taken as the continuum limit of Motzkin and Fredkin groundstates. It thus raises the question of which field theory does?
A possible direction would be to look at the \textit{compact} Lifshitz boson groundstate. As may be observed in \cite{Zhou:2016ykv} in $(2+1)$ dimensions, potential IR divergent terms in the mutual information get canceled against an additional contribution coming from the winding modes present in the compact case. It would thus be interesting to generalize our work for the one-dimensional theory to compact fields.


\begin{acknowledgments}
It is a pleasure to thank Benjamin Doyon for interesting discussions. Ch.B. thanks the Department of National Defense of Canada for financial support to facilitate completion of his PhD. Cl.B. is supported by a CRM-Simons Postdoctoral Fellowship at the Université de Montréal. W.W.-K. was funded by a Discovery Grant from NSERC, a Canada Research Chair, a grant from the Fondation Courtois, and a ``Établissement de nouveaux chercheurs et de nouvelles chercheuses universitaires'' grant from FRQNT.
\end{acknowledgments}

\appendix

\section{Deformation by a singular potential}\label{S:singular}

In this appendix, we return to the original Lifshitz theory~\eqref{E:H_Lifshitz} and perform a nontrivial deformation (for $d\neq 2$) that preserves the Lifshitz scale invariance in all dimensions, as well as the spatial conformal symmetry of the groundstate when $d=1$. Set $\Lambda[\phi]=\int d^d x \, \frac{g}{4}\phi(x)^{2d/(d-2)}$, so that the classical action becomes $S_{\text{cl}}[\phi]=\int d^d x \left(\kappa (\nabla\phi)^2 + \frac{g}{2}\phi^{2d/(d-2)}\right)$, and
\beq\label{E:A_sing}
 A(x) =\frac{1}{\sqrt{2}}\left(\frac{\delta}{\delta\phi(x)}-\kappa \nabla^2\phi +\frac{gd}{2(d-2)}\phi^{\frac{d+2}{d-2}}\right).
\eeq
The corresponding groundstate of $H_{\text{sing}}^{\text{normal}}=\int d^d x \;A^{\dagger}(x)A(x)$ is
\beq\label{E:Psi_sing}
\Psi_0[\phi] = \frac{1}{\sqrt{\mathcal{Z}}}\hspace{1pt}e^{-\frac{1}{2}\int d^d x \left(\kappa (\nabla\phi)^2+\frac{g}{2}\phi^{2d/(d-2)}\right)}\, ,
\eeq
with normalization factor given in \eqref{E:Z_conformal}. The parameter $g$ is dimensionless for all $d$, and the operators $A(x)$ are invariant under Lifshitz rescaling $(\mathbf{x},t)\to(\lambda\mathbf{x},\lambda^2 t)$, as per \eqref{E:dim_phi} and \eqref{E:dim_Pi}.  (The affine field-shift symmetry is lost however.) As a consequence, the groundstate wavefunctional $\Psi_0[\phi]$ is invariant under spatial scaling $\mathbf{x}\to\lambda\mathbf{x}$. Yet, $H_{\text{sing}}^{\text{normal}}$ hardly qualifes as a \textit{bona fide} parent Hamiltonian for $|\Psi_0\rangle$, because it contains the awkward UV-divergent term
\beq\label{E:divterm_sing}
\frac{1}{2}\text{tr}(\kappa\nabla^2  -g\tfrac{d(d+2)}{2(d-2)^2}\phi^{\frac{4}{d-2}})\,,
\eeq
whose physical meaning is rather opaque. Note that this term is nothing but the commutator $\frac{1}{2}\int d^d x \, [ A^{\dagger}(x),A(x)]$. But we can no longer rely on the deformed Lifshitz theory $H_{\text{sing}}^{+}=\frac{1}{2}\int d^d x \; \{A^{\dagger}(x),A(x)\}$, even though it is without divergence, since it differs from $H_{\text{sing}}^{\text{normal}}$ by precisely~\eqref{E:divterm_sing}, which is \textit{not} a multiple of the identity,
so this time $H_{\text{sing}}^{\text{normal}}$ and $H_{\text{sing}}^{+}$ do not share their eigenstates.
 
\subsection*{1.\quad Supersymmetric deformation} 
The way around this difficulty is to deform $H_{\text{sing}}^{+}$ (and its Hilbert space) to include fermionic degrees of freedom \cite{dijkgraaf2010relating}. Define the operators
\beq
Q=\int d^d x \,\bar\psi (x) A(x) \;, \quad\; \bar Q=\int d^d x \,\psi (x)A(x)^{\dagger}\,,
\eeq
with $\{\psi(x),\bar\psi(y)\}=\delta (x-y)$. In the Schrödinger picture, where $\psi(x)$ is a multiplication operator and $\bar\psi(x)=\delta / \delta\psi(x)$, consider the normalized Grassmann-valued functional $\Psi_F[\psi]=\prod_{x}\psi(x)$. Then $\psi(x)|\Psi_F\rangle = 0$ for all $x$. The Hamiltonian
\beq\label{E:H_SUSY_Q}
H^{\text{SUSY}}_{\text{sing}}=\frac{1}{2}\{\bar Q, Q\}
\eeq
is part of a supersymmetric structure
\beq
\{Q, Q\}=0=\{\bar Q, \bar Q\}\,,\quad [Q, H^{\text{SUSY}}_{\text{sing}}]=0=[\bar Q, H^{\text{SUSY}}_{\text{sing}}]\quad
\eeq
with unbroken supersymmetry: for $|\Omega\rangle = |\Psi_0\rangle\otimes |\Psi_F\rangle$, where $\langle \phi | \Psi_0\rangle = \exp (-S_{\text{cl}}[\phi])$ as before, we have
\beq
Q|\Omega\rangle = 0 = \bar Q|\Omega\rangle\,.
\eeq
As in the work of Dijkgraaf \textit{et al}.\ \cite{dijkgraaf2010relating}, the presence of the fermionic sector enables both $Q$ and $\bar Q$ to annihilate $|\Omega\rangle$, making it a groundstate of the \textit{divergenceless} anticommutator $\{\bar Q, Q\}$.
Since the product $\Psi_F[\psi]= \prod_{x}\psi (x)$ has no spatial entanglement, the supersymmetric groundstate $|\Omega\rangle$ and the bosonic groundstate $|\Psi_0\rangle$ possess the same structure of spatial entanglement. Moreover, $H^{\text{SUSY}}_{\text{sing}}$ is an unambiguous parent Hamiltonian for $|\Omega\rangle$. Interestingly, this type of supersymmetric structure naturally emerges when stochastically quantizing a classical action $S_{\text{cl}}$ \cite{parisi1981perturbation}. The parent Hamiltonian $H^{\text{SUSY}}$ is then physically realized as the stochastic field theory whose $t\to\infty$ equilibrium correlations yield the correlations of the quantized action. Expanding~\eqref{E:H_SUSY_Q}, we get
\begin{align}
H^{\text{SUSY}}_{\text{sing}}&=\frac{1}{2}\int d^d x \left[\Pi^2 +\tfrac{1}{2} (S'_{\text{cl}})^2 +\tfrac{1}{2}\big( \bar\psi S_{\text{cl}}'' \psi - \psi S_{\text{cl}}'' \bar\psi\big) \right]\nonumber\\
	&=H^{+}_{\text{sing}} + \frac{1}{4}\int d^d x \left[\bar\psi S_{\text{cl}}''\psi - \psi S_{\text{cl}}''\bar\psi \right].
\end{align}
In the first line, we have written $S_{\text{cl}}^{(n)}=(\delta^n / \delta\phi^n ) S_{\text{cl}}$, for short. We emphasize that $|\Psi_0\rangle\otimes |\Psi_F\rangle$ is a groundstate of the above Hamiltonian, despite the fact that $|\Psi_0\rangle$ is not a groundstate of $H^{+}_{\text{sing}}$. The corresponding Euclidean theory has partition function
\beq\label{E:Z_E}
\mathcal{Z}_E=\int \mathcal{D}\phi\mathcal{D}\psi\mathcal{D}\bar\psi\; e^{-\int d\tau d^d x \;\mathcal{L}_E}\,,
\eeq
where, up to boundary terms,
\beq\label{E:L_E_SUSY}
\mathcal{L}_E=\frac{1}{2}\big(\partial_{\tau}\phi + \tfrac{1}{2}S_{\text{cl}}'\big)^2 - \bar\psi \big(\partial_{\tau} +\tfrac{1}{2}S_{\text{cl}}''\big)\psi\,.
\eeq
Specializing again to $d=1$, we find
\beq\label{E:H_sing_plus}
H_{\text{sing}}^+ = H_{\text{Lif}}+\frac{1}{2}\int dx \left(\frac{g\kappa}{\phi^3}\nabla^2 \phi +\frac{g^2}{4\phi^6}\right),
\eeq
and
\beq\label{E:L_singular}
\begin{aligned}
\mathcal{L}_E	&=\tfrac{1}{2}(\partial_{\tau}\phi)^2 + \tfrac{\kappa^2}{2}(\nabla^2 \phi)^2 +\tfrac{g\kappa}{2}\phi^{-3}\nabla^2 \phi +\tfrac{g^2}{8}\phi^{-6}\\
			&\hspace{10pt}- \bar\psi (\partial_{\tau} - \kappa\nabla^2 + \tfrac{3g}{2}\phi^{-4})\psi\,.
\end{aligned}
\eeq
We will mostly consider the stable case $g>0$, for which the singular potential in~\eqref{E:H_sing_plus} prevents the field from vanishing anywhere. Without loss of generality, we can restrict $\phi$ to strictly positive real values $\phi(x)\in\mathbb{R}^{>0}$. The boundary conditions are chosen to be $\phi(0,t)=\phi(L,t)=h$, for an arbitrary but strictly positive regulator $h=0^{+}$, preventing the divergence of the potential energy term. (Our analysis also holds for negative values of $g$  as long as the regulating condition $\phi\geq h>0$ is maintained.) The normalization factor of $|\Omega\rangle = |\Psi_0\rangle\otimes |\Psi_F\rangle$ is
\beq\label{E:Z_conformal}
\mathcal{Z}=\int_{\phi(0)=h=\phi(L)} \mathcal{D}\phi\; e^{-\int dx \left(\kappa (\partial_x\phi)^2+\frac{g}{2\phi^2}\right)}.
\eeq
One can recognize $\mathcal{Z}$ as the partition function of a single quantum mechanical particle with Euclidean Lagrangian $\mathcal{L}_g=\kappa (\partial_x \phi)^2 +\frac{g}{2\phi^2}$. The theory in spatial dimension \mbox{$d=1$} stands out as having full conformal spatial symmetry. The group of dilations $x\to \lambda x$, $\lambda>0$, may be extended to the $SL(2,\mathbb{R})$ group of transformations
\beq\label{E:SL_group}
x\to x' = \frac{\alpha x + \beta}{\gamma x + \delta} \; , \quad\; \alpha\delta - \beta\gamma = 1\,,
\eeq
with Jacobian determinant
\beq
\frac{\partial x'}{\partial x}=\frac{1}{(\gamma x + \delta)^2}>0 \,.
\eeq
Fields transform as tensor densities of weight $1/2$ corresponding to their length dimension
\beq\label{E:tensor_density_phi}
\phi'(x')=\sqrt{\frac{\partial x'}{\partial x}}\phi(x)=\frac{\phi(x)}{\gamma x + \delta}\,.
\eeq
(The sign of $\gamma x +\delta$ is immaterial, the action being quadratic in the field, so we drop the absolute value bars.) 
The action $\int dx \;\mathcal{L}_g (\phi , \partial_x \phi)$ is called $SL(2,\mathbb{R})$-conformal because it is invariant, up to a boundary term, under the joint transformations \eqref{E:SL_group} and \eqref{E:tensor_density_phi}. The quantized theory, conformal quantum mechanics (CQM), has been known for a long time \cite{dealfaro1976conformal,jackiw1972introducing}. Stochastic quantization of $\mathcal{L}_g$ generates the $(1+1)$-dimensional field theory $H_{\text{sing}}^{\text{SUSY}}$, whose supersymmetric groundstate $|\Psi_0\rangle\otimes|\Psi_F\rangle$ contains the propagators of CQM. Conversely, knowledge of the propagators of CQM is analytic knowledge about $|\Psi_0\rangle\otimes|\Psi_F\rangle$ and its parent theory. Because $|\Psi_F\rangle$ is a spatial product, the bosonic groundstate $|\Psi_0\rangle$ and the supersymmetric groundstate $|\Psi_0\rangle\otimes|\Psi_F\rangle$ have the same spatial entanglements. The (normalized) eigenstates of the CQM Hamiltonian, $H_g = -\frac{1}{4\kappa}\partial_{\phi}^2+\frac{g}{2\phi^2}$, were identified in \cite{jackiw1972introducing,dealfaro1976conformal}:
\beq
(H_g - E)\psi_E =0 \iff \psi_E (\phi) = \sqrt{2\kappa\phi}J_{\nu}(\phi\sqrt{4\kappa E})\,,\;
\eeq
where $\nu=\sqrt{2g\kappa+\frac{1}{4}}$ and $J_{\nu}$ is a Bessel function of the first kind of order $\nu$. We will only consider nonnegative values of $\nu$ (i.e. $g\geq -1/8\kappa$). The propagators are readily computed:
\beq\label{E:prop_nu}
\langle \phi' , x' | \phi , x \rangle = \tfrac{2\kappa\sqrt{\phi' \hspace{1pt}\phi}}{x'-x}\exp \left[{-\kappa\left(\tfrac{\phi'^2 + \phi^2}{x'-x}\right)}\right]I_{\nu}\left(\tfrac{2\kappa \phi' \phi}{x'-x}\right),
\eeq
with $I_{\nu}$ the modified Bessel function of the first kind of order $\nu$. Deep in the bulk, where $\langle\phi' \phi\rangle$ is large compared to $x'-x$ (see below), the singular potential is ineffective and we recover the free massless boson propagator $G_{\mathbb{R}}(\phi' , x' ; \phi , x)=\sqrt{\frac{\kappa}{\pi (x'-x)}}\exp\left[-\frac{\kappa(\phi'-\phi)^2}{x'-x}\right]$. On the other hand, in the limit $g\to 0$ (i.e. $\nu\to 1/2$), and using the fact that $I_{\frac{1}{2}}(z)=\sqrt{\frac{2}{\pi z} }\sinh z$, we find the propagator of the positive-valued free field
\beq\label{E:prop_free_orbifold}
\begin{aligned}
&G_{\mathbb{R}^{>0}}(\phi' , x' ; \phi , x)\\
&\qquad=\sqrt{\frac{\kappa}{\pi (x'-x)}}\bigg(e^{- \tfrac{\kappa(\phi'-\phi)^2}{x'-x}}-e^{-\tfrac{\kappa(\phi'+\phi)^2}{x'-x}}\bigg)\,,
\end{aligned}
\eeq
because nothing is left of the singular potential but the constraint $\phi>0$. We emphasize that the singular deformation in not a perturbation of the original Lifshitz point, but of the positive-valued free field. However, the propagators $G_{\mathbb{R}}$ and $G_{\mathbb{R}^{>0}}$ are indistinguishable deep in the interior. 

\subsection*{2.\quad Correlations in the groundstate}


\noindent Obviously, the field has nonvanishing vacuum expectation values
\beq\label{E:vev_phi_nu}
\langle\phi(x)\rangle_{\nu} = \int_h^{\infty} d\phi \; \phi f_1^{(\nu)}(\phi , x) = \frac{\Gamma(\nu + \tfrac{3}{2})}{\Gamma(\nu +1)}\sqrt{\frac{ x (L-x)}{\kappa L}}\,,
\eeq
where $f_1^{(\nu)}(\phi , x)=\langle h , L |h , 0\rangle^{-1}\langle h , L |\phi , x\rangle\langle \phi , x | h , 0\rangle$ and $h=0^{+}$ is the field regulator introduced below \eqref{E:L_singular} to prevent the divergence of the potential energy term. The regulator $h$ is also necessary to prevent the individual vanishing of the propagators, given by \eqref{E:prop_nu}. However, the integral in~\eqref{E:vev_phi_nu}, and similar integrals,  have a well-defined limit as $h$ is sent to zero. We will work in that limit from now on.
Interestingly, the only difference between first moments with different values of $\nu$ (or $g$) is a real-number prefactor. An even stronger relationship holds between all higher moments (see Appendix \ref{A:moment_nu}~4): 
\beq\label{E:moment_relation}
\langle\phi^{\alpha}(x)\rangle_{\mu}=R(x,x,\mu,\nu)\langle\phi^{\alpha+2(\mu-\nu)}(x)\rangle_{\nu}\,,
\eeq
with
\beq
R(x_1,x_2,\mu,\nu)=\frac{\Gamma (1+\nu)}{\Gamma (1+\mu)}\left(\frac{\kappa L}{ x_1 (L-x_2)}\right)^{\mu - \nu}.
\eeq
In particular, higher moments can be obtained from the \textit{first} moment of the (positive-valued) free boson via the relations $\langle\phi^{1+2\mu}(x)\rangle_{\nu}\propto\langle\phi(x)\rangle_{\nu+\mu}\propto\langle\phi(x)\rangle_{1/2}$. The general expression is
\beq
\langle\phi^{1+2\mu}(x)\rangle_{\nu}=\frac{\Gamma(\mu+\nu+\frac{3}{2})}{\Gamma(\nu+1)}\left(\frac{ x (L-x)}{\kappa L}\right)^{\frac{1}{2}+\mu}.
\eeq
Deep in the bulk, $\langle\phi^{1+2\mu}(x)\rangle_{\nu}\sim (\sqrt{L})^{1+2\mu}$ for any $\nu$.

\begin{figure}
	\includegraphics[scale=1.005]{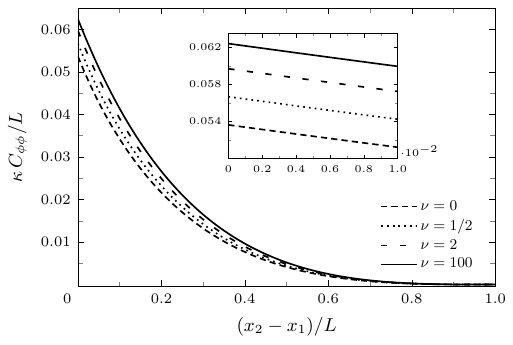}
	\caption{Connected correlator $\frac{\kappa}{L}C_{\phi_1 , \phi_2}(x_1, x_2)$ for a central interval of length $x_2-x_1$ for $\nu=0,1/2, 2, 20$, obtained from \eqref{E:vev_phi_nu} and \eqref{E:phiphi_g_ABC}.  \textbf{Inset:} Small central interval regime. The linear behavior is in perfect agreement with~\eqref{E:connected_phiphi}.}\label{F:phiphi_g}
\end{figure}

\begin{figure}
	\includegraphics[scale=1.007]{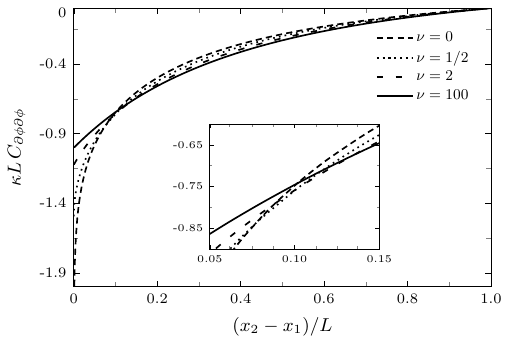}
	\caption{Connected correlator $\kappa LC_{\partial\phi _1, \partial\phi_2}(x_1 , x_2)$  for a central interval of length $x_2-x_1$, and for $\nu=0,1/2, 2, 20$, obtained analytically from differentiating \eqref{E:vev_phi_nu} and \eqref{E:phiphi_g_ABC}.  \textbf{Inset:} Correlations are almost independent of $\nu$ when $x_2 - x_1 \sim 0.1L$. }\label{F:dphidphi_g}
\end{figure}

Once again, the reduced density $\rho_{A\cup B}$ is a separable mixed state for any disconnected subsystems $A,B$ (see Section~\ref{A:separability}). Therefore, $A$ and $B$ are not entangled, and their correlations are non-entangling \cite{ollivier2002quantum, giorda2010gaussian,Adesso:2016ygq,adesso2016introduction}, coming from the statistical mixture left in $\rho_{A\cup B}$, if any. As $A,B$ come into contact, the result does not hold anymore and quantum-driven contact terms are expected. To compute the two-point function $\langle\phi(x_1)\phi(x_2)\rangle_{\nu}$, we find it best to express it in terms of the variable 
\beq
w=\frac{(L_A + L_B)(L_B + L_C)}{L_A L_C}\,,
\eeq
with subinterval lengths $L_A, L_B, L_C$ as in Fig.\,\ref{F:ABA}. We find
\beq\label{E:phiphi_g_ABC}
\begin{aligned}
\langle\phi(x_1)\phi(x_2)\rangle_{\nu} &=\frac{L_B}{\kappa}\left[\frac{\Gamma (\frac{3}{2}+\nu)}{\Gamma (1+\nu)}\right]^2  
\frac{(w-1)^{1+\nu}}{w^{\frac{3}{2}+\nu}}\\
&\quad\times\; \prescript{}{2}{}F_1^{} \left[\tfrac{3}{2}+\nu , \tfrac{3}{2}+\nu ; 1+\nu ; \tfrac{1}{w}\right],
\end{aligned}
\eeq
where $\prescript{}{2}{}F_1^{} (a,b;c;x)$ is the Gaussian hypergeometric function. From \eqref{E:vev_phi_nu} and \eqref{E:phiphi_g_ABC}, we can obtain analytic expressions for the connected correlator $C_{\phi_1,\phi_2}=\langle\phi_1 \phi_2\rangle_{\nu} - \langle\phi_1\rangle_{\nu}\langle\phi_2\rangle_{\nu}$ and, by repeated differentiation, for the connected correlator $C_{\partial\phi_1,\partial\phi_2}=\langle\partial\phi_1 \partial\phi_2\rangle_{\nu} - \langle\partial\phi_1\rangle_{\nu}\langle\partial\phi_2\rangle_{\nu}$. Here we omit the full expressions, but in Fig.\,\ref{F:phiphi_g} we plot $C_{\phi_1 , \phi_2}(x_1 , x_2)$ (in units of $\kappa / L$) for a central interval of length $x_2- x_1$, and different values of $\nu$. In Fig.\,\ref{F:dphidphi_g} we plot $C_{\partial\phi_1 , \partial\phi_2}(x_1 , x_2)$ (in units of $\kappa L$) for the same interval and values. Deep in the bulk, where $w\to 1$, we may expand~\eqref{E:phiphi_g_ABC} in powers of 
\beq\label{E:lambda}
w-1=\frac{L L_B}{L_A L_C}=\frac{L(x_2-x_1)}{x_1 (L-x_2)}\sim \frac{1}{L}\, ,
\eeq
leading to

\beq\label{E:phiphi_g_final}
\langle\phi(x_1)\phi(x_2)\rangle_{\nu} =\frac{(1+\nu)L}{4\kappa } -\frac{x_2 - x_1}{4\kappa}+O(1/L)\,.
\eeq
This expression is not restricted to central intervals. For the special case $\nu=1/2$, we recover
\beq
\langle\phi(x_1)\phi(x_2)\rangle_{1/2} = \frac{3L}{8\kappa } -\frac{x_2 - x_1}{4\kappa}+O(1/L)\,,
\eeq
in agreement with the positive-valued boson. The field connected correlator deep in the bulk is found to be
\beq\label{E:connected_phiphi}
\begin{aligned}
C_{\phi , \phi}(x_1 , x_2)&=\langle\phi(x_1)\phi(x_2)\rangle_{\nu} - \langle\phi(x_1)\rangle_{\nu} \langle\phi(x_2)\rangle_{\nu} \\
				&=\frac{\alpha_{\nu}L}{\kappa}-\frac{x_2 - x_1}{4\kappa} + O(1/L)\,,
\end{aligned}
\eeq 
with $\alpha_{\nu}$ a finite, positive constant independent of $L$:
\beq\label{E:alpha_nu}
\alpha_{\nu}  =\frac{1}{4}\left[1+\nu-\left(\frac{\Gamma(\nu+\frac{3}{2})}{\Gamma(\nu+1)}\right)^2\right].
\eeq
The bulk two-point function~\eqref{E:phiphi_g_final}, and the bulk connected correlator~\eqref{E:connected_phiphi} consist in a constant leading term which diverges like $L$ in the thermodynamic limit, followed by a universal, translation-invariant bulk term linear in $x_2-x_1$, and independent of $\nu$. The universal behavior $\langle\phi(x_1)\phi(x_2)\rangle_{\nu}\sim (x_2 - x_1)^{\Delta}$, with exponent $\Delta=1$, is a joint consequence of the dimension of the fields, $\text{dim}\,\phi=\sqrt{\text{length}}$, and of the absence of a length scale other than $L$. Non-dependence of this term on $\nu$ is yet another indication that the singular potential is ineffective in the bulk.

From~\eqref{E:connected_phiphi} and dimensionality, we expect $\partial\phi\partial\phi$ correlations to vanish in the limit $L\to\infty$. To obtain an analytic bulk expansion for $C_{\partial\phi_1 , \partial\phi_2}$ beyond the leading order we find it necessary to restrict $\nu$ to chosen intervals or values. We provide four cases, beginning with $\nu=0$, and the interval $\nu\in (0,1)$.

\smallskip
\noindent\textbf{Case} $\nu=0$: 
\beq\label{E:dphidphi_v=0}
\begin{aligned}
C_{\partial\phi,\partial\phi}=\frac{\beta_{0}}{\kappa L}&+\frac{1}{4\kappa L}\log \frac{x_2-x_1}{L}
	-\frac{3(x_2-x_1)}{4\kappa L^2}\log\frac{x_2-x_1}{L}\\
	&+\frac{\delta_0}{\kappa L^2}(x_2-x_1)+O(1/L^3)\,,
\end{aligned}
\eeq
with $\beta_0 = -\frac{1}{2}\log 2$, and $\delta_0 = \frac{3}{2}\log 2 -1$.

\smallskip
\noindent\textbf{Case} $0<\nu<1$: 
\beq\label{E:dphidphi_series}
C_{\partial\phi,\partial\phi}=\frac{\beta_{\nu}}{\kappa L}+\frac{\gamma_{\nu}}{\kappa L^{1+\nu}}(x_2-x_1)^{\nu}+O(1/L^2)\,
\eeq
with
\beq\label{E:beta_gamma}
\beta_{\nu}=-1-\frac{1}{4\nu}\;,\quad\; \gamma_{\nu}=4^{\nu}\left[\frac{\Gamma (\nu +3/2)}{\Gamma (\nu +1)}\right]^2 \text{csc }\pi\nu \,.
\eeq
The leading term $\beta_{\nu}/\kappa L$, with coefficient given in \eqref{E:beta_gamma}, is actually valid for all nonzero values of $\nu$. The coefficient of fractional order $\gamma_{\nu}$ diverges near $\nu=0$ and $\nu=1$. As $\nu\to1$, term $O(1/L^2)$ in~\eqref{E:dphidphi_series} can no longer be neglected. We also provide the expansions for the integer values $\nu=1$ and $\nu=2$.

\smallskip
\noindent\textbf{Case} $\nu=1$: 
\beq\label{E:dphidphi_v=1}
\begin{aligned}
C_{\partial\phi,\partial\phi}=\frac{\beta_{1}}{\kappa L}&-\frac{9}{4\kappa L^2}(x_2 - x_1)\log (x_2 - x_1)\\
	&+\frac{\delta_1}{\kappa L^2}(x_2-x_1)+ O(1/L^3)\,.
\end{aligned}
\eeq

\smallskip
\noindent\textbf{Case} $\nu=2$: 
\beq\label{E:dphidphi_v=2}
C_{\partial\phi,\partial\phi}=\frac{\beta_{2}}{\kappa L}+\frac{\delta_2}{\kappa L^2}(x_2-x_1) + O(1/L^3)\,,
\eeq
where $\delta_1 = \frac{9}{2}\log 2-\frac{15}{4}$, and $\delta_2=\frac{45}{8}$. In contrast to the $\phi\phi$ correlator, the non-constant leading term of the $\partial\phi\partial\phi$ correlator is strongly sensitive to the value of $\nu$, not only in its coefficient but even in its functional dependence  to the subinterval length $x_2-x_1$. Case $\nu=0$ presents a quasiconstant term $O(\log (x_2-x_1))$, while case $\nu=1$ has a quasilinear term $O((x_2-x_1)\log (x_2-x_1))$. For general integer $\nu$, it is tempting to expect a quasipolynomial term $O((x_2-x_1)^{\nu}\log (x_2-x_1))$, subleading for $\nu\geq 2$.

Finally, the correspondence \eqref{E:moment_relation} between higher moments can be generalized to multipoint functions deep in the bulk. For the points $0<x_1<x_2<\cdots <x_N<L$ with $x_1\sim L/2 \sim L-x_N$, and writing $\phi(x_i)=\phi_i$, we find
\beq\label{E:multimoment_relation}
\begin{aligned}
\langle&\phi_1^{\alpha_1}\cdots \phi_N^{\alpha_N}\rangle_{\mu}=\\
&\quad R(x_1,x_N,\mu,\nu)\langle\phi_1^{\alpha_1+\mu-\nu}\phi_2^{\alpha_2}\cdots \phi_{N-1}^{\alpha_{N-1}}\phi_N^{\alpha_N+\mu-\nu}\rangle_{\nu}\,.
\end{aligned}
\eeq
These relationships are a manifestation of the fact that bulk propagators are free, see Section 4 of this appendix. The effect of the singular potential on
\eqref{E:multimoment_relation} arises solely from the surface propagators $\langle\phi_{1} , x_{1}|h,0\rangle$ and $\langle h ,L|\phi_N , x_N\rangle$ between the system's boundaries and $\phi_1, \phi_N$. All other propagators are free. More generally, when some points are close to the boundary, more bulk propagators become boundary propagators. If, say, $x_1$ is close to the boundary, both $\langle\phi_{1} , x_{1}|h,0\rangle$ and $\langle\phi_{2} , x_{2}|\phi_{1},x_{1}\rangle$ will be boundary propagators, and $\alpha_2$ will be modified accordingly in~\eqref{E:multimoment_relation}.

\subsection*{3.\quad Entanglement in the groundstate}

We now compute the Rényi entanglement entropies of a boundary interval $A$ as in Fig.\,\ref{F:ABA}, which are given by
\beq
\begin{aligned}
S_n (A)	&=\frac{1}{1-n}\log \int_{h}^{\infty} d\phi  \; \left[f_1^{(\nu)}(\phi, L_A)\right]^n + \frac{1}{2}\log\frac{1}{\epsilon},
\end{aligned}
\eeq
where $f_1^{(\nu)}=\mathcal{Z}^{-1}\langle h , L | \phi , L_A\rangle  \langle \phi , L_A | h , 0\rangle$ is the single-point probability distribution obtained from the propagator \eqref{E:prop_nu}, $h=0^{+}$ is the field regulator introduced below \eqref{E:L_singular}, and $\mathcal{Z}$ is a normalization factor given in \eqref{E:Z_conformal}. Note that the (generalized) Fradkin-Moore formula \eqref{E:S_n_mass_gen1} is not applicable here because its derivation requires that $M$-point functions be Gaussian. In the limit of vanishing regulator, we obtain
\beq\label{E:S(A)}
S_n (A)=\frac{1}{2}\log \frac{L_A(L-L_A)}{\epsilon L}  +b(n,\nu) \,,
\eeq
with
\beq
b(n,\nu)\hspace{-1pt}=\hspace{-1pt}\tfrac{n(1+2\nu)+1}{2(n-1)}\hspace{-1pt}\log\hspace{-1pt} n + \tfrac{1}{1-n}\hspace{-1pt}\log\hspace{-1pt} \tfrac{\Gamma\left((1+n +2n\nu)/2\right)}{\Gamma (1+\nu)^n}-\tfrac{1}{2}\hspace{-1pt}\log4\kappa.
\eeq
Comparing \eqref{E:S(A)} with the entropy \eqref{REmassless1} in the (undeformed) massless case then reveals the same entanglement behavior for all values of $\nu$.  In particular, all $c_n$-functions are identically equal to $1/2$, consistent with the unbroken spatial conformal symmetry of the groundstate wavefunctional deep in the bulk.

\subsection*{4.\quad Multipoint moments}\label{A:moment_nu}

\begin{figure}[t]
\centering
\includegraphics[scale=1.075]{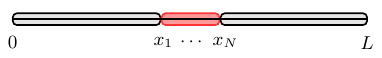}
\caption{$N$ ordered points $x_1<\cdots <x_N$ deep in the bulk.}\label{F:ABC_Npoints}
\end{figure}

Consider $N$ ordered points $x_1<x_2<\cdots <x_N$ deep in the bulk, as shown in Fig.\,\ref{F:ABC_Npoints}, and let us write $\phi(x_i)=\phi_i$. Based on the fact that propagators become free in the bulk, we expect a relationship between the functions 
\beq
\resizebox{1\hsize}{!}{$f_N^{(\nu)}(\pmb{\phi},\mathbf{x})=\frac{\langle h ,L|\phi_N , x_N\rangle}{\langle h ,L|h ,0\rangle} \prod_{i=1}^{N-1}\langle\phi_{i+1} , x_{i+1}|\phi_{i},x_{i}\rangle\langle\phi_1 , x_1|h,0\rangle$}
\eeq
with different values of $\nu$. As in the previous section, $h=0^{+}$ is a field regulator preventing the potential energy term from diverging. The propagators are given in \eqref{E:prop_nu}, which is repeated here for convenience:
\beq
\langle \phi' , x' | \phi , x \rangle = \tfrac{2\kappa\sqrt{\phi' \hspace{1pt}\phi}}{x'-x}\exp \left[{-\kappa\left(\tfrac{\phi'^2 + \phi^2}{x'-x}\right)}\right]I_{\nu}\left(\tfrac{2\kappa \phi' \phi}{x'-x}\right).
\eeq
The only dependence in $\nu$ is in the Bessel function. Assume that $x_1\sim L/2 \sim x_N$.  From \eqref{E:phiphi_g_final}, we know that $\langle \phi_{i+1}\phi_i\rangle \sim L$, and therefore $\langle \phi_{i+1}\phi_i\rangle / (x_{i+1}-x_i)\gg 1$ for each $1\leq i\leq N-2$. Neglecting fluctuations, and with $I_{\nu}(z)\sim e^z / \sqrt{2\pi z}$ for $z\gg 1$, we find that bulk propagators are asymptotic to those of the positive-valued free boson
\beq
\langle\phi_{i+1} , x_{i+1}|\phi_{i},x_{i}\rangle_{\nu}\sim G_{\mathbb{R}^{>0}}(\phi_{i+1} , x_{i+1} ; \phi_i , x_i)\,,
\eeq
as shown in the text in \eqref{E:prop_free_orbifold}. Thus, $f_N^{(\nu)}(\pmb{\phi},\mathbf{x})$ will depend on $\nu$ only through the boundary propagators
\beq\label{E:surface_propagators}
\frac{\langle h ,L|\phi_N , x_N\rangle \langle\phi_1 , x_1|h,0\rangle}{\langle h ,L|h ,0\rangle}\,,
\eeq
and ratios $\rule{0pt}{20pt}f_N^{(\mu)}(\pmb{\phi},\mathbf{x}) / f_N^{(\nu)}(\pmb{\phi},\mathbf{x})$ will be independent of bulk propagators. 
From \eqref{E:vev_phi_nu} we know that $\langle\phi_1\rangle / x_1 \sim L^{-1/2} \sim\langle\phi_N\rangle / (L-x_N)$. Therefore, both $\langle h\phi_1\rangle / x_1$ and $\langle h\phi_N\rangle / (L-x_N)$ will be much smaller than 1 for small values of $h$. With $I_{\nu}(z)\sim \frac{(z/2)^{\nu}}{\Gamma (1+\nu)}$ for $z\ll 1$, we find
\beq
\frac{f_N^{(\mu)}(\pmb{\phi},\mathbf{x})}{f_N^{(\nu)}(\pmb{\phi},\mathbf{x})}=R(x_1,x_N,\mu,\nu)(\phi_1 \phi_N)^{\mu-\nu}\,,
\eeq
with 
\beq
R(x_1,x_N,\mu,\nu)=\frac{\Gamma (1+\nu)}{\Gamma (1+\mu)}\left(\frac{\kappa L}{ x_1 (L-x_N)}\right)^{\mu-\nu}.
\eeq
Thus,
\beq\label{E:multipts_appendix}
\begin{aligned}
&\langle\phi_1^{\alpha_1}\cdots\phi_N^{\alpha_N}\rangle_{\mu}\\
&\;=\int d\pmb{\phi}\; \phi_1^{\alpha_1}\cdots\phi_N^{\alpha_N}f_N^{(\mu)}(\pmb{\phi},\mathbf{x})\\
&\;=R(x_1,x_N,\mu,\nu)\int d\pmb{\phi}\; \phi_1^{\alpha_1}\cdots\phi_N^{\alpha_N}(\phi_1\phi_N)^{\mu-\nu}f_N^{(\nu)}(\pmb{\phi},\mathbf{x})\\
&\;=R(x_1,x_N,\mu,\nu)\langle\phi_1^{\alpha_1+\mu-\nu}\phi_2^{\alpha_2}\cdots \phi_{N-1}^{\alpha_{N-1}}\phi_N^{\alpha_N+\mu-\nu}\rangle_{\nu}\,,
\end{aligned}
\eeq
which is identical to \eqref{E:multimoment_relation}. The single-point relation \eqref{E:moment_relation} is a special case, but its validity is actually slightly more general because it contains no bulk propagator. Thus the above result carries through even if $x$ is close to the boundary, as long as the regulator $h$ is $0^+$. Relationship \eqref{E:multipts_appendix} is easily generalized if some points are close to the boundary. Then \eqref{E:surface_propagators} has to be modified to account for the additional boundary propagators.

\section{Capacity of entanglement} \label{A:capacity}

We investigate another information theoretic quantity, namely the capacity of entanglement (see, e.g., \cite{DeBoer:2018kvc,Kawabata:2021hac,Okuyama:2021ylc} for recent developments). It was introduced in \cite{PhysRevLett.105.080501} as the quantum information analog of the heat capacity for thermal systems. 
The capacity of entanglement for the reduced density matrix $\rho_A$ may be conveniently defined as 
\beq
C_A = \lim_{n\rightarrow1} n^2\partial^2_n\log\text{Tr}\,\rho_A^n =\lim_{n\rightarrow1} n^2\partial^2_n(1-n)S_n(A)\,.
\eeq
It is seen to characterize the width of the eigenvalue spectrum of the reduced density matrix. 

Using expression \eqref{E:EE_2pt_mass} for the Rényi entropy, the capacity of entanglement for the (massive) Lifshitz theory is simply given by
\beq
C_A = \frac{M}2\,.
\eeq
We find that the capacity of entanglement is finite, independent of the mass $m$, follows an area law (the number of surface points $M$ is the one-dimensional analog of a surface  area), and is thus much smaller than the entanglement entropy.
The latter observation is quite different from CFT results \cite{DeBoer:2018kvc} where usually $C_A\sim S_A$.

An interesting take on \cite{DeBoer:2018kvc} is that a capacity of entanglement much smaller than entanglement entropy could be interpreted as entanglement being effectively carried by maximally entangled EPR pairs. Conversely, quantum states for which $C_A\sim S_A$ would be better described by randomly entangled pairs of qubits. 
The capacity of entanglement thus provides insights into the entanglement structure of the groundstates of Lifshitz theories.

For the sake of completeness, 
let us apply the results of Section \ref{S:positive} and Appendix \ref{S:singular} where we study the positive-valued Lifshitz boson and a singular deformation of this theory. For the positive boson, using \eqref{E:EE_pos1} and \eqref{E:EE_pos2}, we find the capacity of entanglement to be
\beq\label{CApos}
C_A^{\rm (pos)}=\frac{M}{2}-\Big(5-\frac{\pi^2}2\Big)\,,
\eeq
where $M=1,2$ corresponds to the number of surface points of $A$. We conjecture this expression to be valid for $M\ge1$.
For the singular deformation we obtain
\beq
C_A^{(\nu)} = (\nu+1/2)^2\psi^{(1)}(\nu+1) -\nu\,,
\eeq
where $\psi^{(n)}(x)$ is the polygamma function of order $n$, and $\nu$ is related to the coupling parameter to the singular potential. This formula was derived for a boundary interval $A$ (i.e. $M=1$). We conjecture that for general subsystems with $M$ surface points 
\beq
C_A^{(\nu)} = \frac{M}{2}-(\nu+1/2)\Big(1-(\nu+1/2)\psi^{(1)}(\nu+1)\Big)\,,
\eeq
such that we recover the positive boson result for $\nu=1/2$ (see Appendix \ref{S:singular}). We note that $C_A^{(\nu)}\le C_A,\; \nu\ge0$.


\providecommand{\href}[2]{#2}\begingroup\raggedright\endgroup

\end{document}